\documentclass[fleqn,usenatbib]{mnras}

\usepackage{epsfig}
\usepackage{times}
\usepackage{amsmath}
\usepackage[british]{babel}             
\usepackage[british]{babel}             
\usepackage{newtxtext}                  
\usepackage[slantedGreek]{newtxmath}    

\usepackage{threeparttable}

\usepackage[T1]{fontenc}                
\usepackage{graphicx}                   
\usepackage{subfig}

\newif\ifAMStwofonts
\newcommand{\target}{PSR\,J1023+0038}

\newcommand{\vsini}{$v_{\rm rot}\,\rm sin\,{\it i}$}

\newcommand{\kms}{\,km\,s$^{-1}$}

\newcommand{\erg}{\,erg\,s$^{-1}$}
\newcommand{\msun}{\,$\rm M_{\sun}$}
\newcommand{\ang}{\,\AA}

\title[Optical spectroscopy of PSR~J1023+0038]
{The binary millisecond pulsar PSR~J1023+0038 -- II. Optical spectroscopy}

\author[T.\,Shahbaz et al. ]
{T.\,Shahbaz,$^{1,2}$\thanks{E-mail: tsh@iac.es}
M.\,Linares,$^{3,4}$         
P.\,Rodr{\'i}guez-Gil$^{1,2}$, 
J.\,Casares$^{1,2}$ \\
$^1$Instituto de Astrof\'\i{}sica de Canarias (IAC), E-38205 La Laguna, 
Tenerife, Spain \\
$^2$Departamento de  Astrof\'\i{}sica, Universidad de La Laguna (ULL), 
E-38206 La Laguna, Tenerife, Spain \\
$^3$Department de F{\'i}sica, EEBE, Universitat Polit{\`e}cnica de 
Catalunya, c/ Eduard Maristany 10, 08019 Barcelona, Spain \\
$^4$Institute of Space Studies of Catalonia (IEEC), E-08034 Barcelona, Spain
}

\pagerange{\pageref{firstpage}--\pageref{lastpage}}
\pubyear{2014}

\begin{document} 

\maketitle 

\begin{abstract} 

\noindent
We present time-resolved optical spectroscopy of the ``redback'' binary 
millisecond pulsar system PSR\,J1023+0038 during both its radio pulsar 
(2009) and accretion disc states (2014 and 2016). 
We provide observational evidence for the companion star being heated during the 
disc-state. We observe a spectral type change along the orbit, from 
$\sim$G5 to $\sim$F6 at the secondary star's superior and inferior 
conjunction, respectively, and find that the corresponding irradiating 
luminosity  can be powered by the high energy accretion 
luminosity or the spin down luminosity of the neutron star. We determine the secondary 
star's radial velocity semi-amplitude from the metallic (primarily Fe and 
Ca) and H$\alpha$ absorption lines during these different states. 
The metallic and H$\alpha$ radial velocity 
semi-amplitude determined from the 2009 pulsar-state observations allows 
us to constrain the secondary star's true radial velocity $K_{\rm 
2}$=276.3$\pm$5.6 \kms\ and the binary mass ratio $q$=0.137$\pm$0.003. By 
comparing the observed metallic and H$\alpha$ absorption-line radial 
velocity semi-amplitudes with model predictions, we can explain the 
observed semi-amplitude changes during the pulsar-state and during the 
pulsar/disc-state transition as being due to different amounts of heating 
and the presence of an accretion disc, respectively.

\end{abstract}


\begin{figure*}
\centering
\includegraphics[width=0.8\linewidth,angle=0]{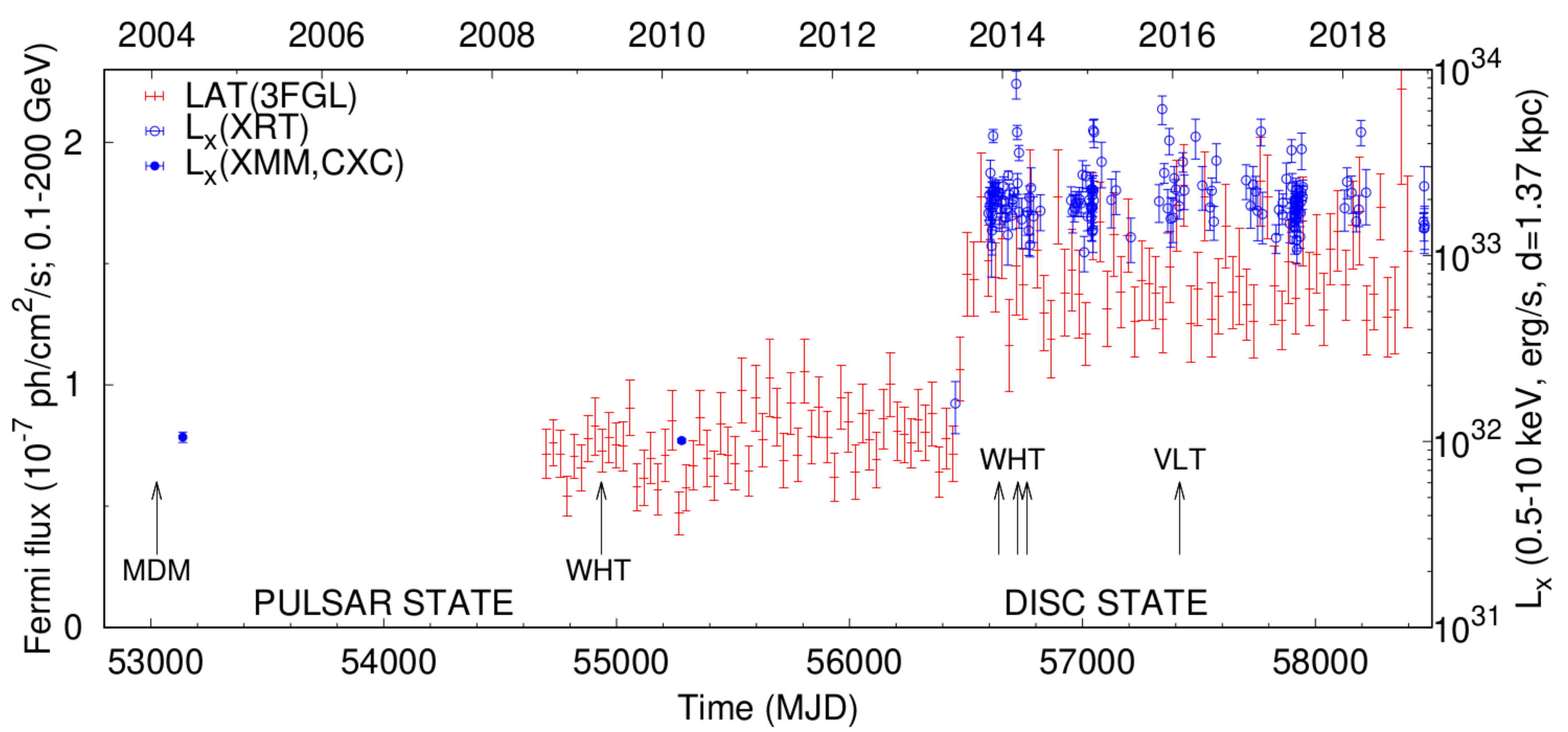}
\caption{
The Fermi LAT light curve is shown with red symbols (photon flux in the 0.1--200 GeV band; from the LAT 3FGL catalogue (\url{https://fermi.gsfc.nasa.gov/ssc/data/access/lat/4yr_catalog/ap_lcs.php)}. The state transition is clearly visible in mid 2013. We also show the 0.5--10 keV X-ray luminosities from Swift-XRT data (open blue circles), as well as the X-ray luminosities measured with  {\it Chandra} \citep{Bogdanov11} and {\it XMM-Newton} \citep{Homer06} before 2013 (filled blue circles). The distance is taken from the known radio parallax \citep{Deller12}. The arrows show the time of the MDM 2004 \citep{TA05}, WHT ISIS 2009, WHT ISIS 2014 and VLT X-SHOOTER 2016 optical observations. 
} 
\label{fig1:lat}
\end{figure*}


\section{Introduction}
\label{sec:intro}

Radio millisecond pulsars (MSP) are fast rotating magnetic neutron stars, 
that were spun up via the transfer of angular momentum from the companion 
star to the neutron star in a low mass X-ray binary (LMXB) \citep{Alpar82, 
Radhakrishnan82}. With the discovery of the first ``transitional'' MSP 
(tMSP) \target\ \citep{Archibald09}, the connection between 
rotation-powered MSPs and accretion-powered LMXBs was established. The 
MSPs M28I \citep{Papitto13} and XSS\,J12270--4859 \citep{Bassa14} have 
also switched back and forth between a rotation-powered radio pulsar state 
(``pulsar-state'') and an accretion-powered state which shows X-ray 
pulsations and accretion disc signatures (``disc-state'').

\target\ was discovered by \citet{Bond02} as part of the ``Faint Images of 
the Radio Sky at Twenty Centimeters'' (FIRST) survey. It was initially 
classified as a cataclysmic variable in 2001 because the optical 
counterpart to the radio source displayed short time-scale flickering and 
a blue optical spectrum with double-peaked emission lines, associated with 
an accretion disc \citep{Szkody03}. Optical photometry taken in 2003 
\citep{Woudt04} and 2004 \citep{Homer06} did not show the rapid, large 
flickering events seen in the 2001 light curve, suggesting that the system 
had changed state. Only a repetitive 4.75476(2)\,h single-humped 
modulation was observed \citep{Woudt04}. Indeed \citet{TA05} confirmed 
this state transition because the optical spectrum taken in 2003 was 
dominated by strong absorption features and lacked the prominent emission 
lines observed in the 2001 spectrum. They performed a time-resolved 
optical spectroscopic and photometric study of \target, in the what we 
know now to be the 'pulsar-state'. They found the companion star to be a 
late-type G5 star with an absorption-line radial velocity semi-amplitude 
268$\pm$3\kms modulated on an orbital period $P_{\rm orb}$=4.754\,h. The 
optical light curves taken in 2004 revealed a single-humped modulation, 
explained in terms of an X-ray heated companion star in a system with an 
orbital inclination angle of $i\sim 55^{\circ}$.  Combining the 
photometric and radial velocity studies led to the conclusion that the 
system was not a cataclysmic variable with a white dwarf, but instead an 
X-ray binary harbouring a neutron star. Indeed, the X-ray binary scenario 
also explained the 2004 X-ray observations which were dominated by a hard 
X-ray power-law component \citep{Homer06}.

In 2007 the system was detected as a radio pulsar with a spin period of 
1.69\,ms \citep{Archibald09}, confirming that the primary was indeed a 
neutron star and that \target\ is a low-mass X-ray binary. This was the 
first direct evidence of an MSP in a binary system transitioning between 
two distinct states, and gave significant support to the ``recycled'' 
scenario for the origin of MSPs. Between 2008 and 2012 \target\ was 
consistently observed as an eclipsing radio millisecond pulsar, where the 
radio eclipses are attributed to ionized material being forced off the 
companion star by the pulsar wind, as seen in other similar MSPs called 
"black-widows" \citep{Fruchter88}.

In 2013 June, after about 10 years in the pulsar-state, \target\ 
transitioned back to the disc-state, as witnessed by the increased X-ray 
and gamma-ray flux (Fig.\,\ref{fig1:lat}) and the disappearance of radio 
pulsations. \target\ was detected as a radio pulsar in 2013 June 15, but 
eight days later it was not detected in the radio \citep{Stappers14}. 
After the transition, the system brightened by $\sim$\,1\,mag in the 
optical and showed several broad double-peaked emission lines 
\citep{Halpern13, Linares14}. There was also a disappearance of radio 
pulsations and an increase in the X-ray and gamma-ray luminosities by a 
factor of $\sim$20 and $\sim$10, respectively \citep{Linares14b,Takata14,Stappers14, 
Patruno14}.  All these factors provided compelling evidence that an 
accretion disc had re-formed, indicating that \target\ had switched back 
from a radio millisecond pulsar (pulsar-state) to a low-mass X-ray binary 
(disc-state).

In \citet[][paper I]{Shahbaz15} we presented the results of our optical 
photometric campaign, which revealed unprecedented fast variability. 
Here 
we present the results of our optical spectroscopic campaign of \target\ 
in the disc-state, including those summarized in \citet{Linares14}, 
as well as archival optical spectroscopy taken during the pulsar-state.
Previous spectroscopic studies do not cover full orbital cycles,
have poor spectral resolution or 
do not compare the different states \citep{TA05,Wang09, Hakala18}.
We perform a spectral type, radial velocity curve analysis and compute 
Doppler maps of the H$\alpha$ emission-line in both the pulsar- and 
disc-state.
Finally, we compare the 
radial velocity semi-amplitude with model predictions, which account for 
the effects of heating and an accretion disc.

\begin{figure*}
\centering
\includegraphics[width=1.0\linewidth]{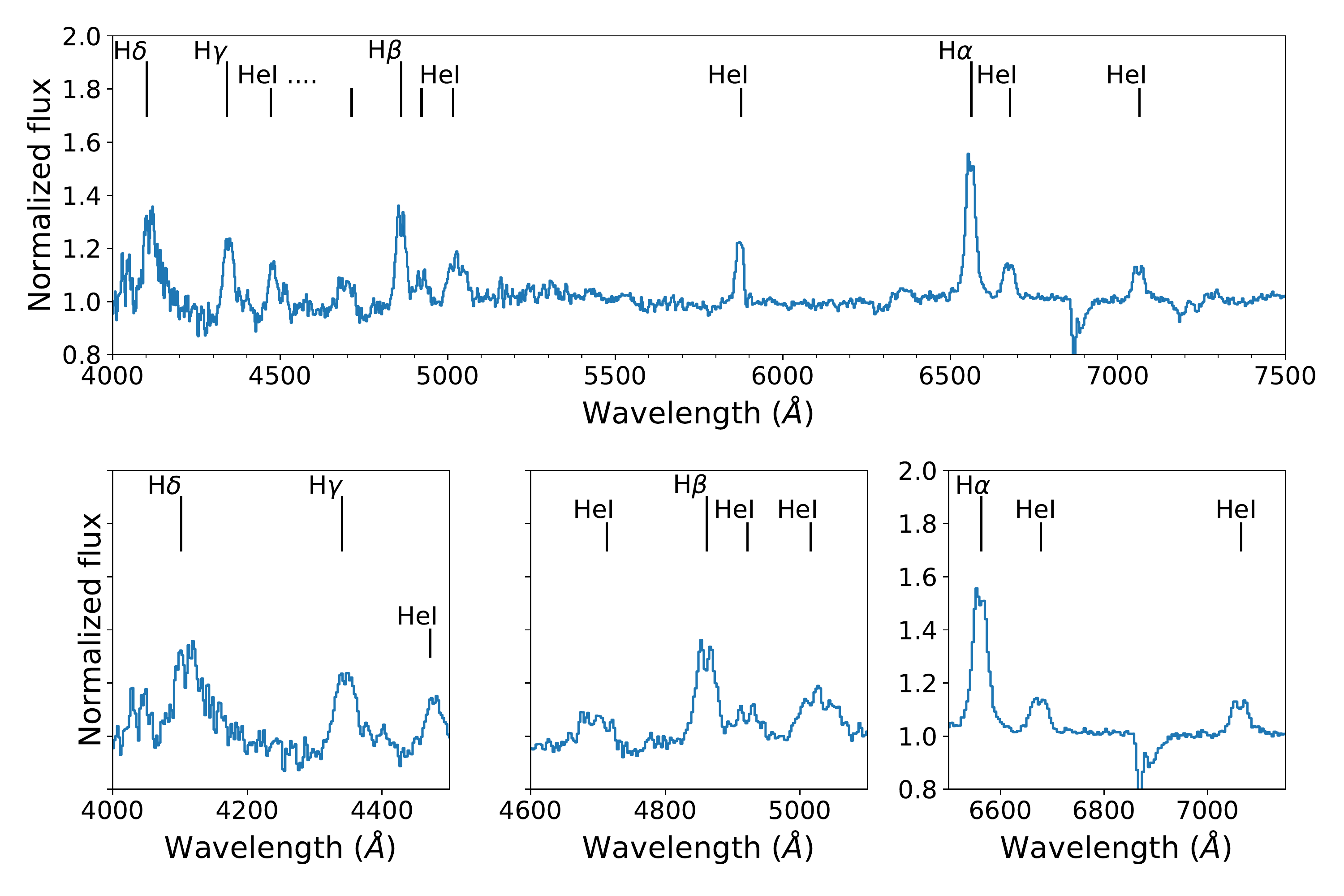}
\caption{Normalized phase-averaged 
ACAM spectrum of \target\ in the disc-state. 
The main emission lines are identified (vertical lines).
}
\label{fig2:spectrum_average}
\end{figure*}

\begin{table*}
\footnotesize
\centering
\caption{Summary of optical spectroscopic observations of \target.}
\centering
\begin{tabular}{l l l c c c c c c c}
\hline
Telescope  & Instrument      & Range             & Disp. (\AA/pixel) & UT date & UT time & Exposures & S/N per  & Orbital & Airmass  \\
(diameter) & Configuration & ($\lambda$, \AA) & FWHM Res. (km/s)       & start   & start--stop   &        & spectrum & phase   & \\
\hline
\\
\multicolumn{10}{c}{--- Observations presented in this paper ---} \\
\\
WHT (4.2-m)   & ACAM   & 3500--9400 & 3.4 & 2013-12-15  & 02:34--06:42  & 22$\times$600\,s &  $\sim$11 & 0.88--1.75 & 1.13--1.73 \\
              & V400+0.5\arcsec & & 350  &  &  &  &  &   \\
WHT (4.2-m)   & ISIS & B:3700--5300 & 0.88 & 2014-03-06 & 22:02--02:43  & 27$\times$600\,s &  $\sim$11 & 0.84--1.82 & 1.13--1.43 \\
              & R600B+1.0\arcsec & & 100   &  &  &  &  &   \\
              & ISIS &  R:5500--7200  & 0.98 & 2014-03-06 & 22:02--02:43  & 27$\times$600\,s &  $\sim$19 & 0.84--1.82 & 1.13--1.43 \\
              & R600R+1.0\arcsec & & 70   &  &  &  &  &   \\
WHT (4.2-m)   & ACAM  & 3500--9400 & 3.4 & 2014-04-14 & 20:15--01:05  & 28$\times$600\,s & $\sim$19 & 0.32--1.34  & 1.13--1.65 \\
              & V400+0.5\arcsec &   & 290 & &  &  &  &   \\ 
\\
\multicolumn{10}{c}{--- Observations downloaded from archives  ---} \\
\\
WHT (4.2-m)  & ISIS    & B:4300--4700 & 0.22 & 2009-04-12 & 20:13--01:13  & 26$\times$600\,s &  $\sim$3 & 0.53--1.50 & 1.07--1.62 \\
             & H2400B+0.82\arcsec &  & 20   &  &  &  &  &   \\
             & ISIS  & R:6200--6900 & 0.50 & 2009-04-12 & 20:13--01:13  & 26$\times$600\,s &  $\sim$10 & 0.53--1.50 & 1.07--1.62 \\
             & R1200R+0.82\arcsec &  & 24   &  &  &  &  &   \\
VLT (8.2-m)  & X-SHOOTER  & VIS:5300--10200 & 0.2 & 2016-01-30 & 04:02--08:28  & 42$\times$372\,s & $\sim$50 & 0.12--1.97  & 1.11--1.36 \\
             & ECHELLLE+0.9\arcsec &    & 40 & &  &  &  &   \\ 
             & X-SHOOTER  & UBV:3000--5560 & 0.2 & 2016-01-30 & 04:02--08:28  & 42$\times$372\,s & $\sim$50 & 0.12--1.97  & 1.11--1.36 \\
             & ECHELLLE+0.9\arcsec&   & 70 & &  &  &  &   \\ 
\hline
\end{tabular}
\label{table:log}
\end{table*}

\section{OBSERVATIONS AND DATA REDUCTION}
\label{sec:obs}

\subsection{WHT observations}

We observed \target\ between November 2013 and April 2014 with the William 
Herschel Telescope (WHT), at the Roque de los Muchachos Observatory on La 
Palma, in order to obtain low-resolution optical spectra of the system in 
the disc state \citep{Linares14}. We used the double-armed Intermediate 
dispersion Spectrograph and Imaging System (ISIS) optical spectrometer 
with the R600B and the R600R gratings as well as the Auxiliary-port CAMera 
(ACAM) with the 400 lines/mm transmission volume phase holographic (VPH) 
grating. These observations were taken about 6 months after \target\ 
transitioned from the radio pulsar to the disc state around 2013 June 30 
\citep{Patruno14,Stappers14}.

The images were first de-biased, and flat-fielded using standard 
procedures within \textsc{iraf}\footnote{\textsc{iraf} is distributed by 
the National Optical Astronomy Observatory, which is operated by the 
Association of Universities for Research in Astronomy, Inc., under 
cooperative agreement with the National Science Foundation. 
\url{http://iraf.noao.edu/}}. After subtracting the bias level, the images 
were divided by a median sky flat field that was normalised by fitting 
high order spline functions to remove the detector specific spectral 
response. The 1-D spectra were then extracted using optimal extraction 
\citep{Horne86} as implemented in \textsc{pamela} \citep{Marsh89}, which 
is part of the 
\textsc{starlink}\footnote{\url{http://starlink.eao.hawaii.edu/starlink}} 
software. The wavelength calibration was done using CuNe/CuAr arc lamp 
spectra taken before, after and between science spectra, extracted from 
the same region as the closest target spectrum in time. Sets of about 20 
to 50 line centroid positions, depending on the instrument, were fitted 
with a 4--6 order polynomial to obtain the dispersion relation, which was 
interpolated linearly in time to account for instrumental flexure when 
calibrating the spectra. The wavelength calibration was subsequently 
checked and refined using the sky emission lines, which were also used to 
measure the spectral resolution (full width at half-maximum, FWHM, of the 
sky lines). The exposure time, wavelength range, dispersion and resolution 
are shown in Table~\ref{table:log}.

\subsection{Archival data}

We downloaded the Isaac Newton Group archival data of \target\ carried out 
on 2009 April 12 with ISIS on the WHT (P.I. Jonker). ISIS was fitted with 
the H2400B and the R1200R gratings in the blue and red arm, respectively 
and a 0.82 arcsec slit width was used. The spectra were extracted using 
the same procedure as the WHT 2014 data. The exposure time, wavelength 
range, dispersion and resolution are shown in Table~\ref{table:log}.

We also downloaded the ESO Very Large Telescope (VLT) archival data of 
\target\ taken on 2016 January 30 under programme 096.D-0808 (P.I. Bassa) 
with the X-SHOOTER medium resolution (R= 4000 to 7000) spectrograph 
\citep{Vernet11}. A slit width of 1.0 arcsec in the UVB arm 
(2989--5560\ang) and 0.9 arcsec in the VIS arm (5337--10200\ang) and NIR 
arms (0.994--2.478\,$\mu$m) was used. The UVB and VIS arm CCDs were binned 
by a factor of 2 and the exposure times were 360\,s in the UVB arm, 372\,s 
in the VIS arm and 400\,s in the NIR arm. A total of 42 spectra were taken 
in nodding mode in each arm. The data reduction pipeline was used to 
optimally extract and calibrate the spectra \citep{Freudling13}. The mean 
dispersion was 10\kms and the spectral resolution measured from the sky 
lines was 40\kms. In this paper we only use the UVB and VIS arm spectra, 
since no telluric stars were observed. The exposure time, wavelength 
range, dispersion and resolution are shown in Table~\ref{table:log}.

\section{Average spectrum}
\label{sec:average}

In Fig.\,\ref{fig2:spectrum_average} we show the disc-state normalized 
phase-averaged WHT ACAM 2014 spectrum of \target. Broad Balmer and He 
lines emission lines are clearly seen, which also show a double-horned 
profile related to the presence of an accretion disc.


\begin{figure}
\centering
\includegraphics[width=1.05\linewidth]{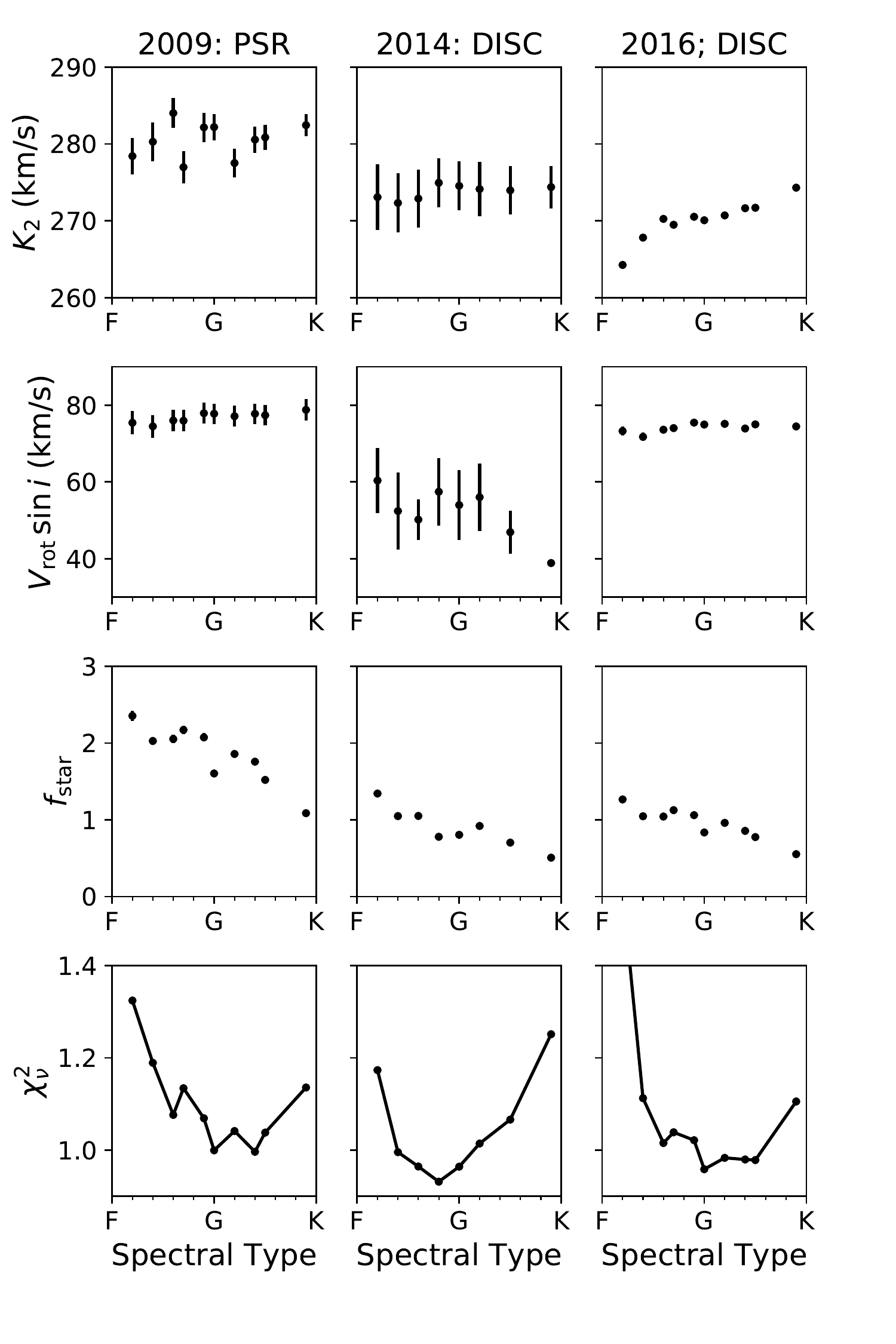}
\caption{
The cross-correlation radial velocity analysis and the optimal subtraction 
to determine the spectral type of the secondary star. The results are 
shown using all orbital phases. As a function of template star spectral 
type, we show the values of $K_{\rm 2}$ and determined from a circular 
orbit fit to the radial velocities, and the values for \vsini, $f_{\rm 
star}$ and $\chi^2$ obtained from the optimal subtraction. The left, 
middle and right columns are the results for the ISIS 2009 (pulsar-state), 
ISIS 2014 (disc-state) and X-SHOOTER 2016 (disc-state) spectra, 
respectively.
}
\label{fig3:optsub}
\end{figure}

\begin{figure*}
\hspace{0mm}
\includegraphics[width=1.05\linewidth]{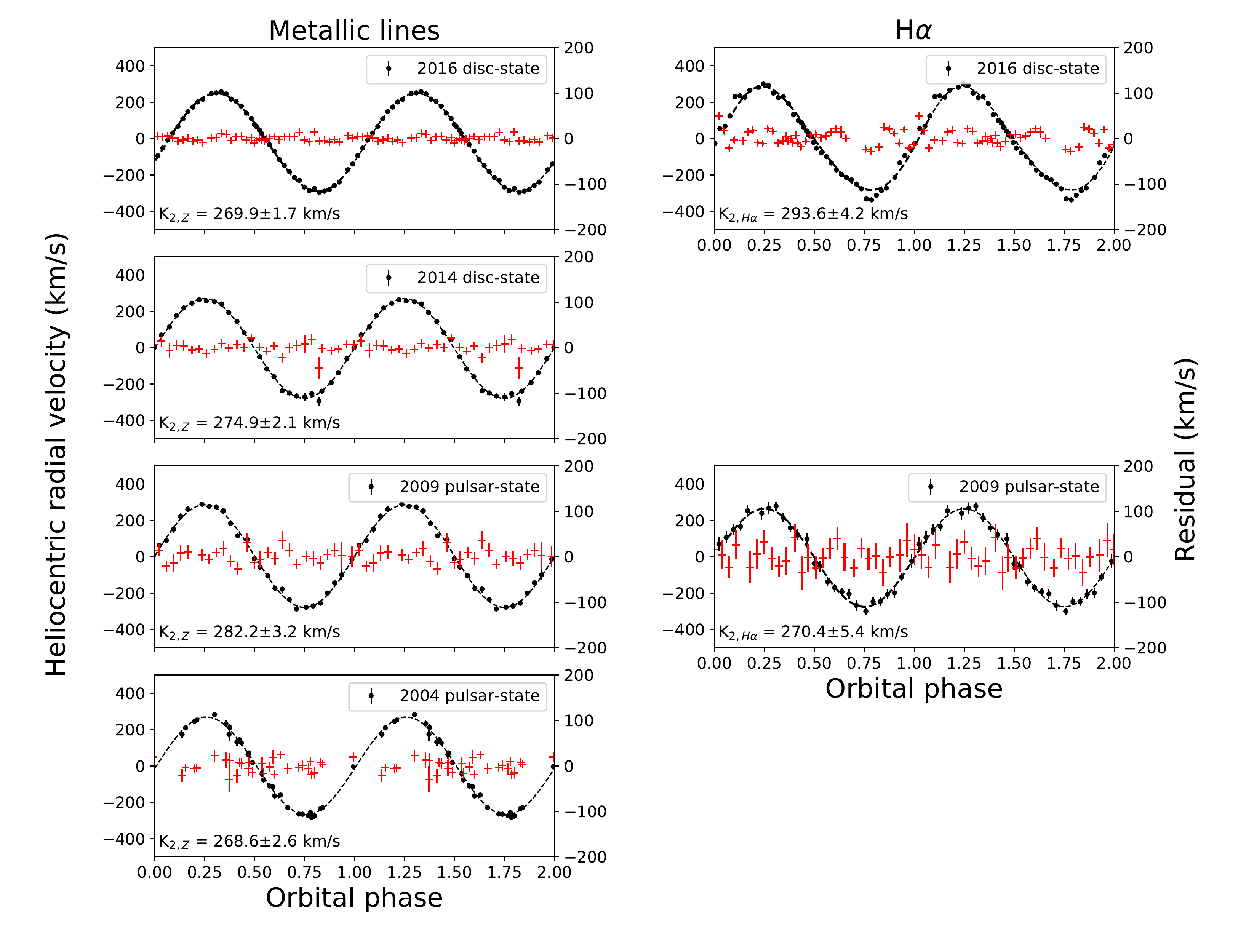}
\caption{
Left: The heliocentric radial-velocity curve (filled circles) of the 
secondary star in \target, determined using the metallic (primarily Fe and 
Ca) lines. We show the 2004 pulsar-state \citep{TA05}, ISIS 2009 
pulsar-state, ISIS 2014 disc-state and the X-SHOOTER 2016 disc-state radial 
velocities. Right: The heliocentric radial-velocity curve (filled circles) 
of the secondary star in \target, determined using the residual H$\alpha$ 
absorption-line. We show the ISIS 2009 pulsar-state and the X-SHOOTER 2016 
disc-state radial velocities. The solid line shows an eccentric orbit fit 
to the data and the red crosses show the residual after subtracting the 
fit. The data have been folded on the orbital ephemeris and are shown 
twice for clarity.
}  
\label{fig4:fig_rv}
\end{figure*}

\begin{figure*}
\centering
\includegraphics[width=1.0\linewidth]{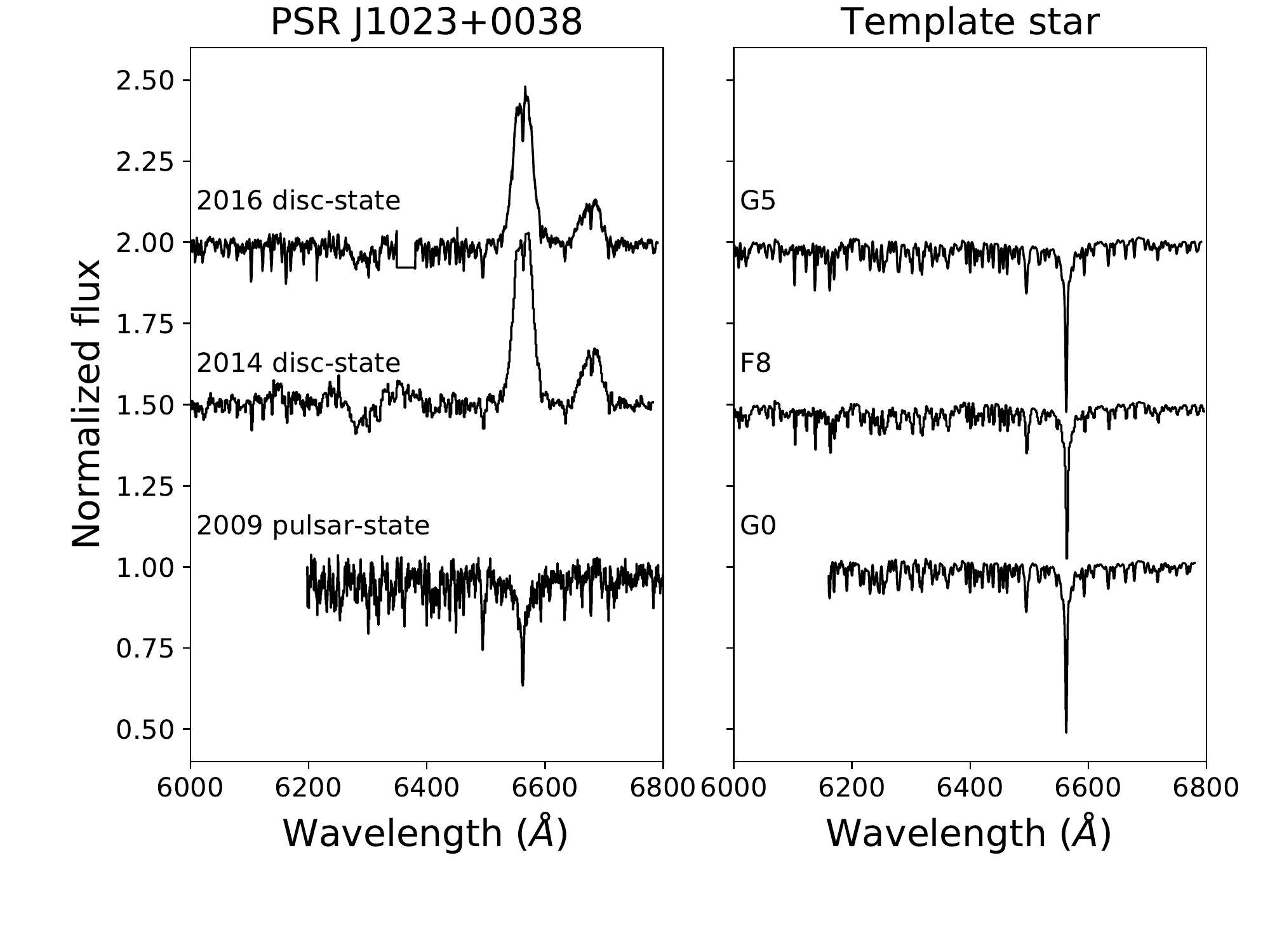}
\caption{
Left: the X-SHOOTER 2016 spectrum taken in the disc-state, 
the ISIS 2014 spectrum taken in the disc-state and the ISIS 2009 spectrum 
taken in the pulsar-state. (The gap in the X-SHOOTER 2016 data is due to an 
artifact in the reduction pipeline). Right: the corresponding optimally 
broadened  template star  spectrum. The spectra have been 
offset for clarity.
}
\label{fig5:fig_spectrum_vsini} 
\end{figure*}

\begin{figure}
\includegraphics[width=1.0\linewidth]{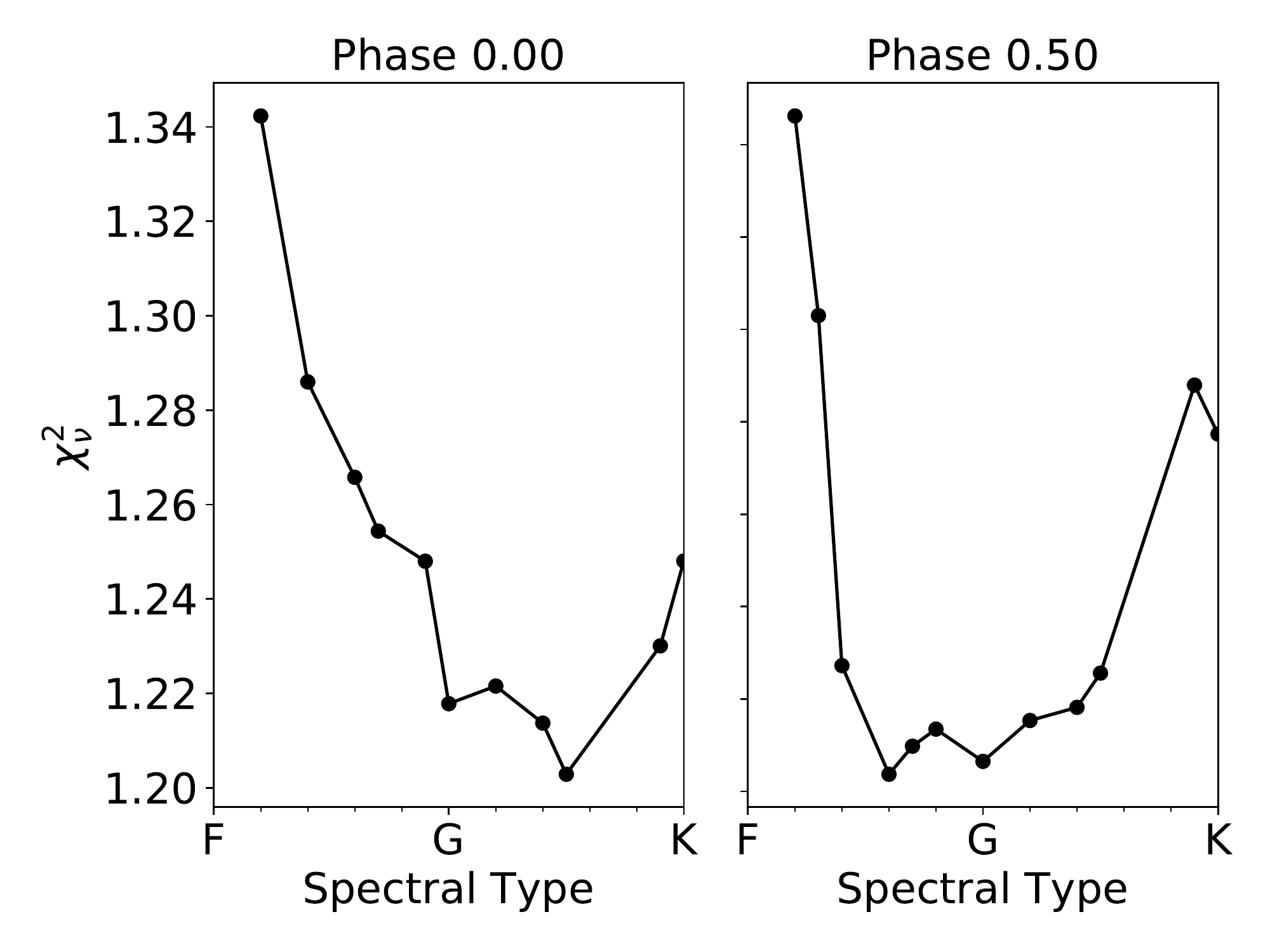}
\caption{
The results of the optimal subtraction to determine the spectral type of 
the secondary star using spectra taken around phase 0.0 and 0.5. We find 
the spectral type to be $\sim$G5 and $\sim$F5, around phase 0.0 and 0.5, 
respectively.
}
\label{fig6:fig_orbit_sptype}
\end{figure}

\begin{table*}
\caption{
Results of the fit to the radial velocity data using the metallic 
and H$\alpha$ absorption-line, taken at different epochs. Note that in each case, 
spectra between orbital phase  0.0 and 1.0 have been used.
}
\begin{minipage}{\textwidth}
\centering
\begin{tabular}{l c c c c }
\hline 
Parameter              &   2004
                                           &    2009           &    2014          &   2016       \\ \hline
State & pulsar & pulsar & disc & disc \\
\\
\multicolumn{5}{c}{--- Metallic absorption lines  --- } \\
\\
$\gamma$  (\kms)       &    0.1$\pm$2.7    &    3.5$\pm$3.0    &    -4.1$\pm$2.0    &  5.2$\pm$0.9  \\
$K_{\rm 2,Z}$ (\kms)   &  268.6$\pm$3.5    &  282.2$\pm$3.2    &  274.9$\pm$2.1   &  270.0$\pm$1.7  \\
$T_{\rm 0}$
\footnote{HJD 2450000+} (d) & 3025.0007$\pm$0.0005  & 4934.6491$\pm$0.0047   
& 6723.0635$\pm$0.0027    & 7417.7871$\pm$0.00015 \\
\vsini\ (\kms)         & -            &   77.7$\pm$2.7    &     -         &   74.9$\pm$0.9  \\
$f_{\rm star}$         & -            &    1.6$\pm$0.5      &    0.78$\pm$0.3    &    0.84$\pm$0.01   \\
Spectral Type          &   G5--G7\footnote{Taken from \citet{TA05}.}     &   G0--G4       &     F6--G0    &   F8--G2      \\
\\
\multicolumn{5}{c}{--- H$\alpha$ absorption-line  --- } \\
\\

$\delta \phi$ (\kms)         & - &   0.08$\pm$0.01      & - &   0.019$\pm$0.008  \\
$\gamma$ (\kms)              & - &   -5.5$\pm$4.6       & - &  -14.8$\pm$3.0  \\
$K_{\rm 2,H\alpha}$ (\kms)   & - &  270.4$\pm$5.4       & - &   293.6$\pm$4.2  \\
$e$                          & - &   0.02$\pm$0.01      & - &   0.08$\pm$0.02 \\
\hline
\end{tabular}
\end{minipage}
\label{table:rv}
\end{table*}

\section{Analysis}
\label{sec:analysis}

In order to measure the absorption-line radial velocity curve and 
determine the spectral type of the companion star in \target, we use 
cross-correlation and optimal subtraction techniques, respectively 
\citep[see][for details]{Marsh94}. All analysis was performed with the 
\textsc{molly}\footnote{\url{http://deneb.astro.warwick.ac.uk/phsaap/software/}} 
package. For a given template star we optimally subtract the 
variance-weighted Doppler-shifted target spectrum from the rotationally 
broadened template star spectrum. We perform a $\chi^2$ test on the 
residuals of the subtraction, where the optimal value for the spectral 
type is obtained by minimizing $\chi^2$. Since both procedures require 
template star spectra, we compiled a library of high signal-to-noise A to 
K star template spectra with luminosity class V from the UVES Paranal 
Observatory Project (UVESPOP) archive \citep{Bagnulo03}. We normalize the 
spectra by determining a low-order spline fit to the continuum level and 
then divide the spectra by the fitted continuum. We degrade the template 
star spectra to match the same spectral resolution and heliocentric 
velocity scale as the target and interpolate all the spectra onto a 
logarithmic wavelength scale with the pixel size set by the target 
spectra.

\subsection{Radial velocity curve}

Firstly, we measure the projected orbital velocity of the companion star 
in \target\ by cross-correlating the spectra of the target with the 
template star of interest. We normalize the spectra similar as above and 
interpolate the spectra onto a logarithmic wavelength scale set by the 
target spectra. As the absorption lines in the target spectrum are 
rotationally broadened, the template star is broadened before performing 
the cross-correlation analysis, using the broadening estimated by 
\cite{MC15b}. We use the wavelength range 5500--6900\ang\ which contains 
numerous weak metallic absorption lines of Fe, Ca and Mg, excluding the H 
and He emission lines and interstellar features. We perform a 
least-squares fit of the radial velocities versus time using a circular 
orbit. The orbital period $P_{\rm orb}$ was fixed at the pulsar timing 
value $P_{\rm orb}$=0.1980963569(3)\,d \citep{Archibald13}. $T_{\rm 0}$ 
was constrained to lie near the middle of the time span over which the 
observations were taken, and defines orbital phase 0 as inferior 
conjunction of the secondary star (see results in Table\,\ref{table:rv}).

\citet{TA05} determine the radial velocities of \target\ in 2004 in the 
pulsar-state by  cross-correlating spectra of \target\ with template stars.
The wavelength range  and spectral type template stars  they use 
in the cross-correlation covers the wavelength range and template stars 
star used here. Here we use their radial velocities and perform the same 
least-squares fit as above.

\subsection{Spectral type}

Using the orbital parameters derived from the radial velocity fits above, 
we Doppler-correct and average the target spectra to the reference frame 
of the template star of interest, over a defined orbital phase range.  In 
principle, the absorption features arising from the companion star should 
be relatively sharp because the Doppler shifts due to orbital motion have 
been corrected. However, the absorption lines are still broadened, due to 
the instrumental resolution of the spectrograph, the rotation of the 
secondary star and by changes in the orbital velocity during individual 
exposures. The broadening function that affects the target's absorption 
lines consists of the convolution of the instrumental profile with the 
rotational broadening profile of the secondary star. There is also a 
further smearing due to the orbital velocity shift of the secondary star 
during a given exposure $\rm \Delta t$ at a given orbital phase, $\rm 
\Delta \rm V_{\rm smear} \sim | \frac{\partial V_{\rm 2}}{\partial t}
|\Delta \rm t_\phi$. The orbital smearing is maximal at the conjunction phases 
and for an exposure time of 600\,s it is $\sim$60\kms. Therefore, it is 
necessary to correct for smearing because the smearing is comparable to the
rotational broadening \citep{MC15b}. For each target exposure, we first smear 
the template spectrum by convolution with a rectangular profile with an amount corresponding to the target's radial velocity and exposure time, and then 
compute the average template star spectrum.

To determine the secondary star's spectral type we compare a broadened 
version of the average, smeared template star spectrum with the target 
spectrum. We broaden the template star from 50 to 100 \kms\ by convolution 
with the \citet{Gray92} rotational profile. We subtract a constant, 
$f_{\rm star}$, representing the fraction of light from the template star, 
multiplied by a smeared, rotationally broadened version of the template 
star.  We eliminate long-scale trends in the residual spectrum by applying 
a high-pass Gaussian filter of FWHM=400\kms. We use a linear 
limb-darkening coefficient of 0.62 \citep{Al-Naimiy78} appropriate for 
6500\ang\ and a mid-GV star. We perform a $\chi^2$ test on the high-pass 
filtered residuals of the subtraction, where the optimal values of \vsini\ 
and $f_{\rm star}$ were obtained by minimizing $\chi^2$. For the optimal 
subtraction procedure we use the wavelength range 6000--6900\ang\, which 
is common to all data sets, excluding the emission lines. We also exclude 
the region 6250--6350\ang\ because it is contaminated by the weak Doppler 
smeared diffuse interstellar bands at 6283\ang\ and the 6300\ang\ [OI] sky 
emission-line.

\begin{figure*}
\includegraphics[width=1.0\linewidth]{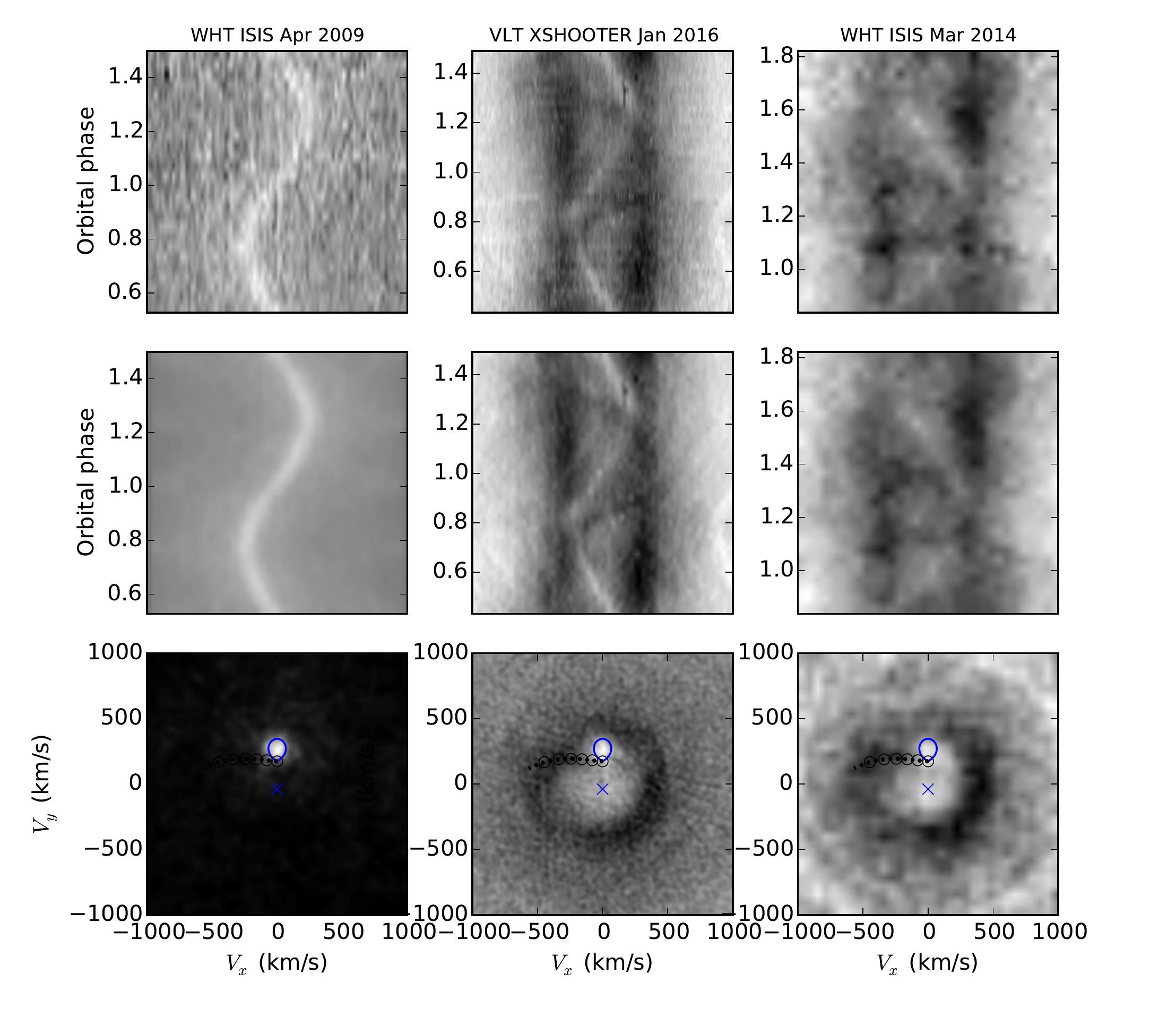}
\caption{
Trailed spectra and Doppler maps of the H$\alpha$ emission-line at 
different epochs. From left to right, WHT ISIS 2009, VLT X-SHOOTER 2016 
and WHT ISIS 2014. From top to bottom; trailed spectra, computed spectra 
and Doppler maps. 
White and black show absorption and emission, respectively.
The ISIS 2009 pulsar-state spectra clearly show the 
narrow H$\alpha$ absorption-line from the secondary star (white S-wave), which can also 
be seen in the VLT X-SHOOTER 2016 and WHT ISIS 2014 disc-state spectra. 
The Roche lobe of the secondary star is also plotted for $q$=0.141 and 
$K_{\rm 2}$=270\kms. The circles show the gas stream and the blue cross, 
the position of the neutron star.
}
\label{fig7:fig_dmaps}
\end{figure*}

\begin{figure}
\centering
\includegraphics[width=1.0\linewidth]{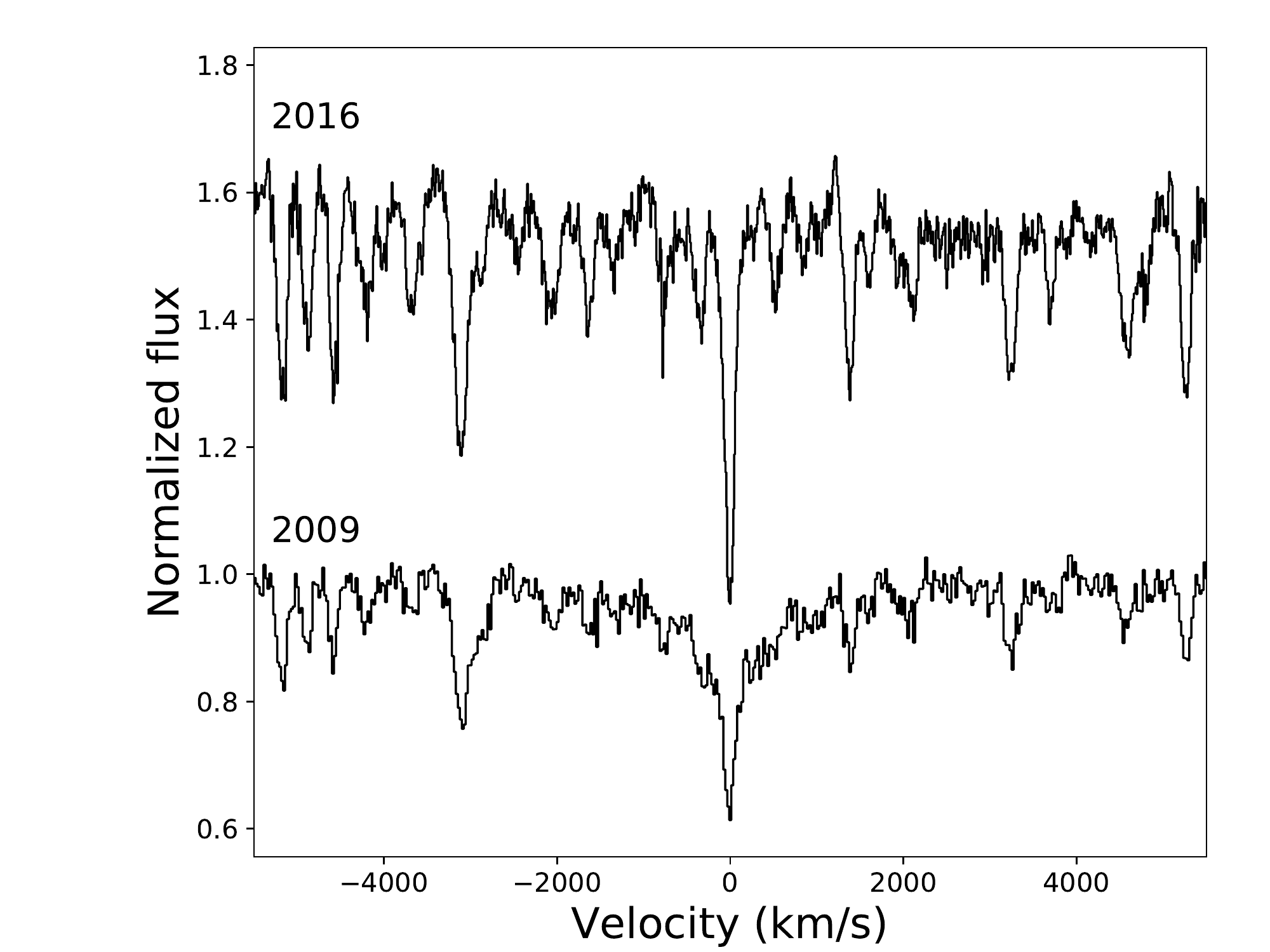}
\caption{
Isolating the narrow H$\alpha$ narrow absorption component arising from the 
secondary star in \target. We show the average H$\alpha$ profile in the 
rest frame of the system's centre-of-mass (top) and the average spectrum of the 
secondary star without the accretion disc profile, showing the narrow 
H$\alpha$ absorption component (bottom). 
The spectra have been offset for clarity.
}  
\label{fig8:fig_ha_profile}
\end{figure}


\begin{figure}
\centering
\subfloat[Diagnostic diagram.]{\includegraphics[width=90mm]{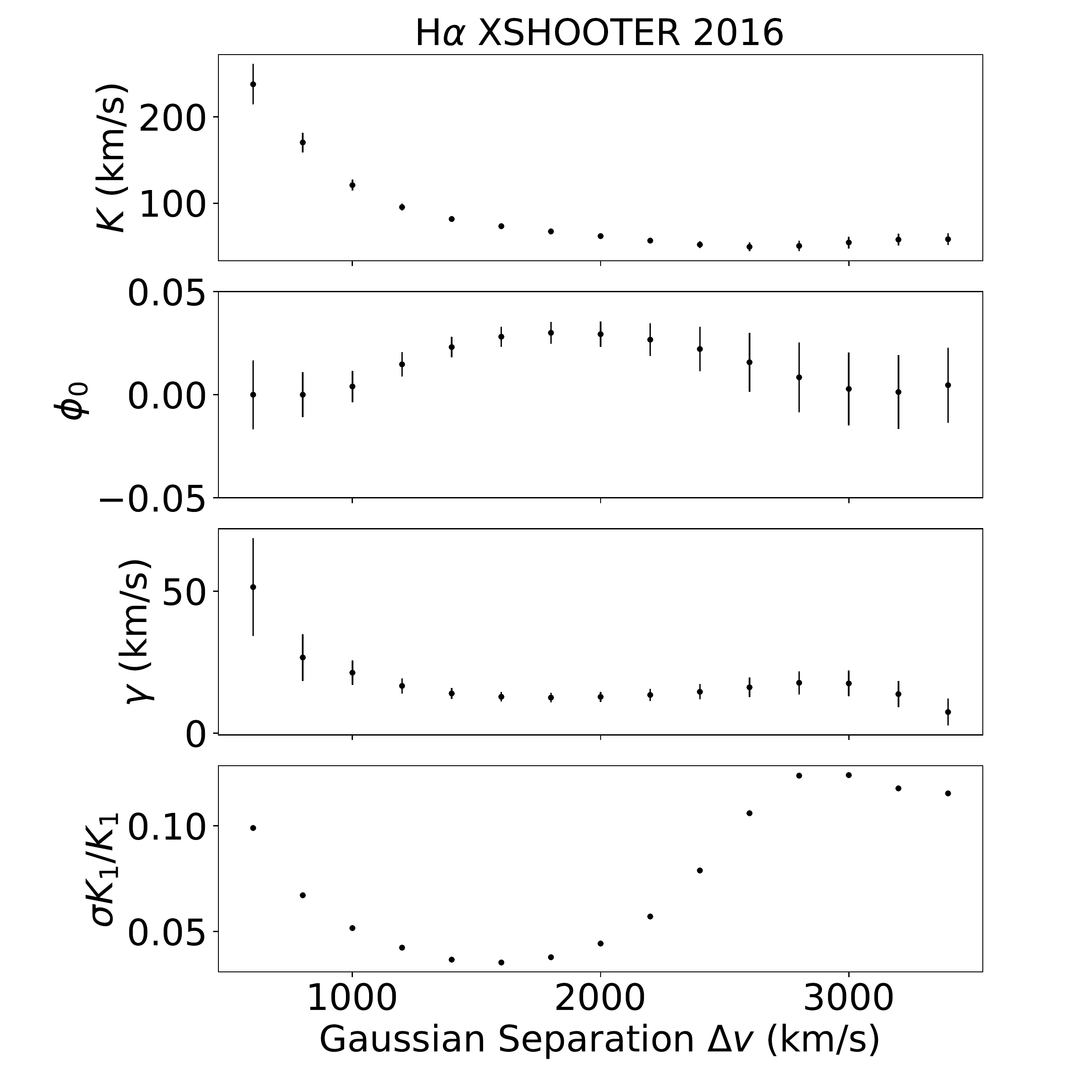}
\label{fig9:fig_diagnostic_a}}

\subfloat[Emission-line radial velocity curves.]{\includegraphics[width=90mm]{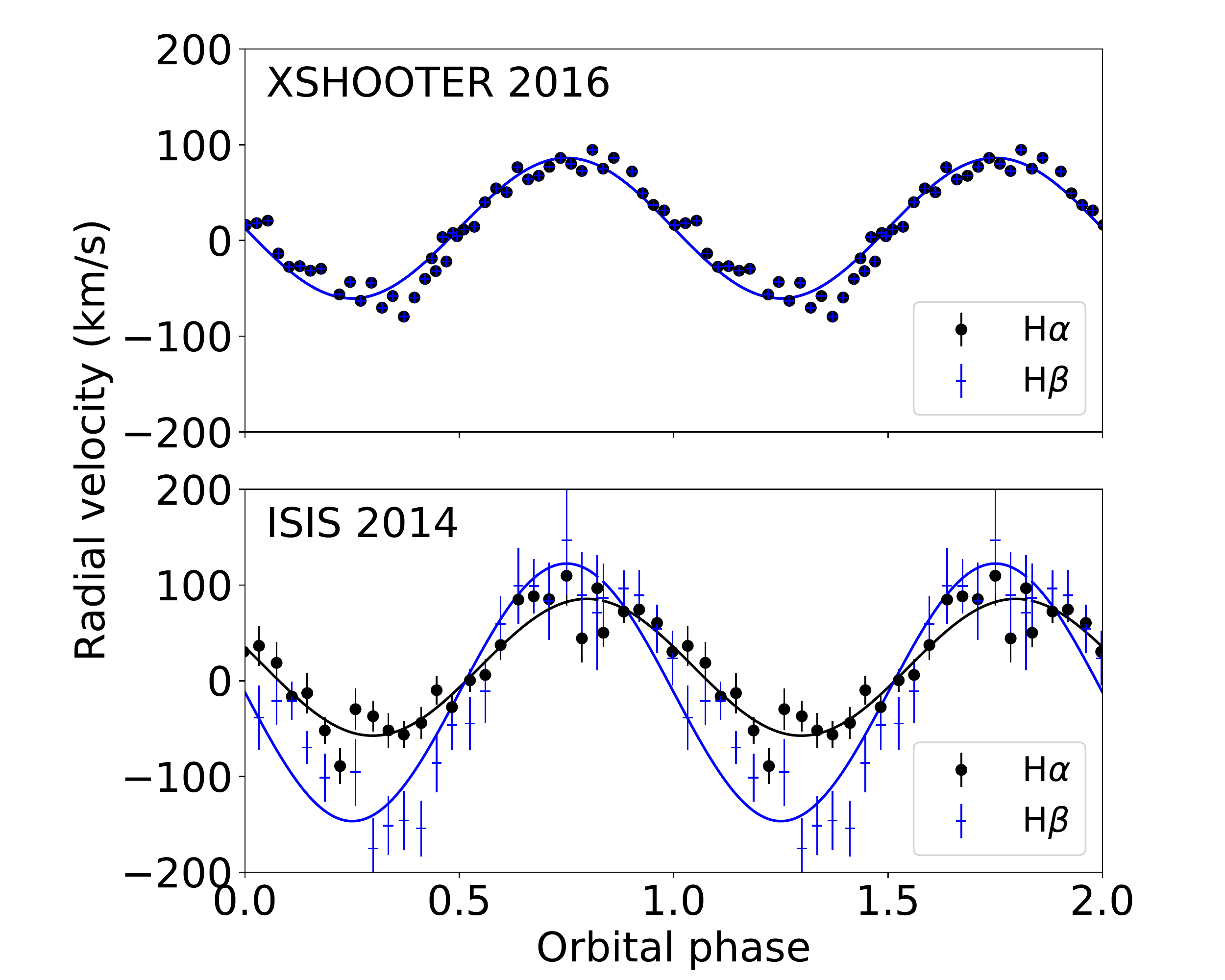}
\label{fig9:fig_diagnostic_b}}
\caption{
(a) An example of the diagnostic diagram for the H$\alpha$ emission line 
of the X-SHOOTER 2016 The panels show the fitted sine-wave parameters are 
shown as a function of the Gaussian separation $\rm \Delta v$. (b) The 
optimal result of the double-Gaussian technique to measure the H$\alpha$ 
(black) and H$\beta$ (blue) emission-line radial velocity curve. The 
resulting best-fit sinusoidal model is shown (solid line). We show the 
results for the X-SHOOTER 2016 (top) and ISIS 2014 (bottom) disc-state 
spectra.
}
\end{figure}


\section{Results}
\label{sec:results}

\subsection{Metallic absorption lines}


In Fig.\,\ref{fig3:optsub} we show the results using the ISIS 2009, ISIS 
2014 and X-SHOOTER 2016 spectra of \target\ between orbital phase 0.0 and 
1.0, for template stars in the F to K spectral range. The $\chi^2$ versus 
spectral type obtained from the optimal subtraction has a minimum at 
G2$\pm$2, F8$\pm$2 and G0$\pm$2 for the ISIS 2009 pulsar-state, ISIS 2014 
disc-state and X-SHOOTER 2016 disc-state, respectively. The uncertainties 
in the spectral type were estimated from a parabolic fit to the minimum. 
For each data set, the result of an circular orbit fit to the radial 
velocities obtained using the optimal spectral type is shown in 
Table\,\ref{table:rv}. For the optimal spectral type, we obtain metallic 
absorption-line radial velocity semi-amplitudes of 
$K_{\rm 2,Z}$=282.2$\pm$3.2\kms, 274.9$\pm$2.1\kms\ and 270.0$\pm$1.7\kms, 
respectively. The 1$\sigma$ errors are quoted where the error bars have 
been rescaled so that the reduced $\chi^2$ of the optimal subtraction fit 
is 1.  Fig.\,\ref{fig4:fig_rv} (left) we show the radial velocity curves 
obtained from the ISIS 2009, ISIS 2014 and X-SHOOTER 2016 spectra folded 
on the ephemeris determined from the optimal spectral type. 
The value for $T_0$ determined independently for each dataset is 
consistent with the calculated $T_0$ adopting an 
ephemeris with constant orbital period.


To determine the secondary star's rotational broadening the spectral 
resolution of the data should be less than the combined effects of the 
instrumental resolution plus orbital smearing.  The ISIS 2014 spectra has 
the worst resolution and hence cannot be used to determine the secondary 
star's rotational broadening. The instrumental spectral resolution is 24, 
70 and 40\kms\ for the ISIS 2009, 2014 and X-SHOOTER 2016 spectra, 
respectively. The maximum smearing is $\sim$60\kms\ for the ISIS 2009 and 
2014 spectra, and $\sim$35\kms\ for the X-SHOOTER 2016 spectra.  For the 
ISIS 2009 and X-SHOOTER 2016 spectra we obtain a \vsini\ of 
77.7$\pm$2.7\kms\ and 74.9$\pm$0.9\kms, respectively. The 1$\sigma$ errors 
are quoted where the error bars have been rescaled so that the reduced 
$\chi^2$ of the optimal subtraction fit is 1. However, the ISIS 2014 data 
with the worst resolution, cannot resolve the star's rotational broadening 
and hence the \vsini\ we obtain is biased towards the blended lines. In 
Fig.\,\ref{fig5:fig_spectrum_vsini} we show the Doppler-averaged spectrum 
of the ISIS 2004, ISIS 2009 and X-SHOOTER 2016 data taken as well as the 
optimal rotationally broadened template star for each data set. As one can 
see, the ISIS 2009 spectrum shows deeper absorption lines compared to the 
template star, indeed for any spectral type $f_{\rm star}$ is greater than 
1, suggesting that the secondary star is peculiar (see 
Section\,\ref{sec:discussion}).


\subsection{Effects of irradiation}


Although the binary has a circular orbit, the observed radial-velocity 
curve of the secondary star can appear non-circular. Heating effects can 
distort the radial-velocity curve leading to an apparent eccentric orbit 
\citep{Davey92,Shahbaz00}. We therefore fit the X-SHOOTER 2016 
radial-velocity curve with an eccentric orbit to determine the effects of 
heating. We perform a least-squares fit of the radial velocities versus 
time using an eccentric orbit of the form,

\begin{equation}
V_{\rm 2}=  \gamma + K_{\rm 2}[\cos(\theta+\omega) + e\cos\omega],
\end{equation}

\noindent
where $\gamma$ is the systemic velocity of the binary, $\theta$ is the 
true anomaly that varies with time, $e$ is the eccentricity of the orbit, 
$\omega$ is the periastron angle and $K_{\rm 2}$ is the radial velocity 
semi-amplitude \citep{Iglesias-Marzoa15}. The value for $\theta(t)$ is 
given by $\tan[\theta(t)/2] = \sqrt{\frac{1+e}{1-e}} \tan(E/2)$, where $E$ 
is the eccentric anomaly, obtained by numerically solving the
Kepler equation $E-e \sin E = \frac{2\pi}{P}(t-T_{\rm 0})$, where 
$T_{\rm 0}$ is the time of periastron passage and $P_{\rm orb}$ is the 
orbital period of the system. For the X-SHOOTER 2016 spectra, the G0 
template star gives a fit with $e$=0.017$\pm$0.002, which is significantly 
better than a circular orbit at the 99.4 per cent level.

To see if the spectral type and thus effective temperature is different on 
the heated and non-heated hemisphere of the secondary star, we repeat the 
spectra type determination analysis in Section\,\ref{sec:analysis} by 
averaging the target spectra around orbital phase $\phi$=0.00$\pm$0.15 and 
orbital phase $\phi$=0.50$\pm$0.15. For the X-SHOOTER 2016 spectra, with 
high signal-to-noise, we find that the secondary star's spectral type 
around phase 0.0 and 0.5 is significantly different, $\sim$G5 and 
$\sim$F6, respectively (see Fig.\,\ref{fig6:fig_orbit_sptype}).

\subsection{H$\alpha$ absorption-line}

A trailed spectrogram of the ISIS 2009 pulsar-state data clearly shows 
narrow H$\alpha$ in absorption (see Fig.\,\ref{fig7:fig_dmaps}), which 
arises from the secondary star. For the disc-state spectra, the accretion 
double-peaked emission-line profile contaminates the narrow H$\alpha$ 
absorption-line. Despite this, narrow H$\alpha$ absorption is clearly seen 
in the data taken with the highest spectral resolution. It is clearly 
observed in the X-SHOOTER 2016 spectra and to lesser extent in the ISIS 
2014 spectra, which has a poorer spectral resolution. To determine the 
radial velocities of the H$\alpha$ absorption-line, we cross-correlate 
each individual target spectrum with a rotationally broadened H$\alpha$ 
absorption-line profile, using only the region around H$\alpha$. We then 
perform a least-squares fit of the radial velocities versus orbital phase 
using an eccentric orbit and determine the radial velocity semi-amplitude 
$K_{\rm 2,H\alpha}$ (see Section\,\ref{sec:analysis}). 
For the ISIS 2014 and X-SHOOTER 2016 disc-state spectra, before we perform the 
cross-correlation, we isolate the narrow H$\alpha$ absorption-line by 
subtracting the average accretion disc profile. The narrow H$\alpha$ 
absorption-line is not clearly resolved in the ISIS 2014 spectra, mainly 
due to the spectral resolution of the data, and so we only analyse the 
X-SHOOTER 2016 spectra. In Fig.\,\ref{fig4:fig_rv} (right) we show the 
radial velocity curves obtained from the ISIS 2009 and X-SHOOTER 2016 
spectra, where we measure $K_{\rm 2,H\alpha}$=270.4$\pm$5.4 and 
293.6$\pm$4.2\kms, respectively (the 1$\sigma$ errors are quoted where the 
error bars have been rescaled so that the reduced $\chi^2$ of the fit is 
1). 
The ISIS 2009 radial velocity curve is consistent with a circular orbit, however, the 
X-SHOOTER 2016 radial velocity curve is eccentric. It is not clear if this 
eccentricity is real or if it is due to the uncertainty in disentangling the narrow H$\alpha$ absorption-line  modulation from the double-peaked disc emission-line profile.
In Fig.\,\ref{fig8:fig_ha_profile} the average Doppler-corrected H$\alpha$ 
absorption-line spectrum of the X-SHOOTER 2016 and ISIS 2014 spectra in 
the rest frame of the secondary star are shown.



\section{Doppler maps}
\label{sec:maps}

Doppler tomography is used to deduce the accretion structures in binary 
systems \citep[see][for a review]{Marsh01}. The method inverts 
phase-resolved spectra into an equivalent image of brightness distribution 
in velocity space \citep{Marsh88}. It is able to separate various sources 
of emission, such as from the secondary star or accretion disc, in 
velocity space. Some of the basic assumptions of Doppler tomography are 
that velocity vectors rotate with the binary, motion is parallel to the 
orbital plane and the flux from any point is constant in time. Violations 
of these assumptions do not imply that Doppler tomography cannot be 
performed, but care must be taken when interpreting the Doppler maps.

To compute the Doppler maps of \target\ we used the normalised continuum 
subtracted spectra and the Python/C++ maximum entropy Doppler tomography 
code \footnote{\url{https://github.com/trmrsh/trm-doppler}} developed by 
T. Marsh. We model both absorption and emission-line modulated components. 
In Fig.\,\ref{fig7:fig_dmaps} we show the trailed spectra and Doppler maps 
of the H$\alpha$ emission-line seen in the 2014 and 2016 disc-state. We 
also compute the Doppler map of the ISIS 2009 pulsar-state spectra, which 
shows the narrow H$\alpha$ absorption-line arising from the secondary 
star. We do not compute the Doppler maps of the H$\beta$ and other lines 
because the ISIS 2009 and 2014 spectra do not have sufficient 
signal-to-noise for a comparison with the X-SHOOOTER spectra. The 
disc-state trailed spectra show the characteristic double-peaked 
tramlines, a signature of a Keplerian accretion disc, that show up as a 
constant emission ring-like structure of high intensity in the Doppler 
maps. The trailed spectra also show a narrow absorption, which is in phase 
with the secondary star, and a narrow emission-line feature which is 
shifted by $\sim$-0.1 phase with respect to the secondary star, which 
arises from the accretion disc. Indeed, the crossing of this emission-line 
feature with the H$\alpha$ absorption-line feature at orbital phase 
$\sim$0.7 seems to be responsible for the larger deviations observed in 
the H$\alpha$ velocities (see Fig.\,\ref{fig4:fig_rv}).The relative 
strength of the blue and red emission-line peaks vary over the orbital 
period. and results in an axisymmetric Doppler map with enhanced emission 
between orbital phase 0.25 and 0.50.

\subsection{The neutron star's radial velocity}
\label{sec:diagnostic}

The emission lines generated from a uniform symmetric accretion disc 
around a neutron star will have the same velocity shift as the neutron 
star. The centroids of the lines will produce a sinusoidal phase-dependent 
modulation corresponding to the orbital motion of the neutron star. The 
addition of other sources of phase-dependent modulation, such as from a 
bright spot, results in a radial velocity curve that is offset in phase 
with respect to the true motion of the neutron star. Therefore, in order 
to exclude these contributions, one examines the wings of the 
emission-line profile corresponding to material orbiting at small disc 
radii from the neutron star, where the contamination by other sources are 
assumed to be minimal.


To measure the radial velocities of the Balmer emission lines we use the 
well established double-Gaussian technique of \citet{Schneider80}. This 
technique allows one to extract radial velocity curves from the wings of 
the emission-line profile, which are expected to follow the motion of the 
neutron star. By convolving the emission-line with double-Gaussian 
function with varying separation the radial velocity curve is determined. 
We apply this technique to the high spectral resolution ISIS 2014 and the 
X-SHOOTER 2016 spectra, where we can clearly see the double-peaked 
emission-line profile from the accretion disc. We use a Gaussian width of 
600\kms\ and vary the separation $\rm \Delta v$ from 600 to 3600\kms\ in 
steps of 200\kms. At each Gaussian separation the resulting radial 
velocity curve was fitted with a sine function of the form

\begin{equation}
V =  \gamma - K\sin[2\pi(\phi-\phi_0)]
\end{equation}

\noindent
where $V$ is the radial velocity, $K$ the velocity semi-amplitude, 
$\gamma$ the systemic velocity, $\phi$ the orbital phase, and $\phi_0$ is 
the phase at superior conjunction of the neutron star. The results of the 
radial velocity analysis as well as the fractional error in the amplitude 
($\sigma K/K$), which is a function of the Gaussian separation, are 
inspected in the form of a diagnostic diagram \citep{Shafter86}. By 
plotting such a diagram it is possible to select the value of $K$ that 
most closely represents the actual $K_{\rm 1}$. The point beyond which the 
noise in the continuum begins to dominate the signal from the 
emission-line wings corresponds to a sharp increase in $\sigma K/K$. 
Usually $K_{\rm 1}$ obtained from a diagnostic diagram is the value 
corresponding to when $\sigma K/K$ is at a minimum (see 
Fig.\,\ref{fig9:fig_diagnostic_a}). For the H$\alpha$ emission-line we 
obtain $ K_{\rm 1}$=72$\pm$5\kms\ and $\phi_0$=0.05$\pm$0.01, whereas for 
the H$\beta$ emission-line we obtain $K_{\rm 1}$=135$\pm$9\kms\ and 
$\phi_0$=0.05$\pm$0.01 for the ISIS 2014 spectra (see 
Fig.\,\ref{fig9:fig_diagnostic_b}). For the X-SHOOTER 2016 spectra we 
obtain $K_{\rm 1}$=73$\pm$3\kms\ and $\phi_0$=0.028$\pm$0.005 using the 
H$\alpha$ emission-line and $K_{\rm 1}$=118$\pm$6\kms and 
$\phi_0$=0.050$\pm$0.007 using the H$\beta$ emission-line. Note that the 
H$\alpha$ and H$\beta$ $K_{\rm 1}$ values for the 2014 and 2016 spectra 
are consistent with each other.

\section{DISCUSSION}
\label{sec:discussion}

\subsection{The neutron star's radial velocity}


The projected semi-major axis of the pulsar orbit measured from radio 
timing observations, $x_{\rm 1}= a_{\rm 1}\sin\,i$=0.343356 light-seconds 
\citep{Jaodand16} allows one to determine the radial velocity 
semi-amplitude of the neutron star 
$K_{\rm 1} = 2\pi c x_{\rm 1} /P_{\rm orb}$ = 38\kms\ (where $a$ is the 
binary separation and $c$ is the speed of light). The fact that we obtain 
$K_{\rm 1}$ values  (see Section\,\ref{sec:diagnostic}) 
that are different for the H$\alpha$ and H$\beta$ 
emission lines wings with relatively large phase offsets, which do not 
agree with the expected value derived from radio timing, means that the 
wings of the emission-line profiles do not follow the motion of the 
neutron star. 
Indeed, it has been shown that 
bright-spot non-Keplerian motion can induce a measured $K_{\rm 1}$ larger 
than the true value \citep{Smak70,Paczynski68}.
The motion of the emission lines reflect the region where 
that emission lines are excited, further out in the accretion disc.

\subsection{Doppler maps}

The X-ray and optical flaring mode-switching behaviour of \target\ in the 
disc-state has been extensively studied by a number of authors 
\citep{Linares14b,Tendulkar14,Bogdanov15,Shahbaz15}. 
The optical and X-ray passive and active mode-switching behaviour have transition 
time-scales of <20\,s, and it has been suggested that they are due to fast of a propeller, where matter in the disk in-flow is propelled away by
the rapidly rotating neutron star magnetosphere \citep{Campana16}. 

\citet{Hakala18} were able to isolate the passive and active-state spectra 
from time resolved H$\alpha$ spectroscopy taken in 2017.  Narrow H$\alpha$ 
absorption-line feature can be clearly seen in the active-state spectra, 
which is not present in the passive-state spectra. They also found 
evidence for a lack of H$\alpha$ emission around orbital phase 0.25--0.50 
in the active-state and interpret it as matter being ejected from the 
system via the propeller effect \citep{Wynn97}. Our 2014 and 2016 
observations also shows narrow H$\alpha$ absorption-line feature but in 
contrast to \citet{Hakala18} we observe excess H$\alpha$ emission around 
orbital phase 0.25--0.50.  Note that \citet{Hakala18} use the orbital 
ephemeris given in \citet{Jaodand16}, which agrees with our ephemeris (see 
Table\,\ref{table:rv}).

\begin{figure}
\centering
\includegraphics[width=1.0\linewidth]{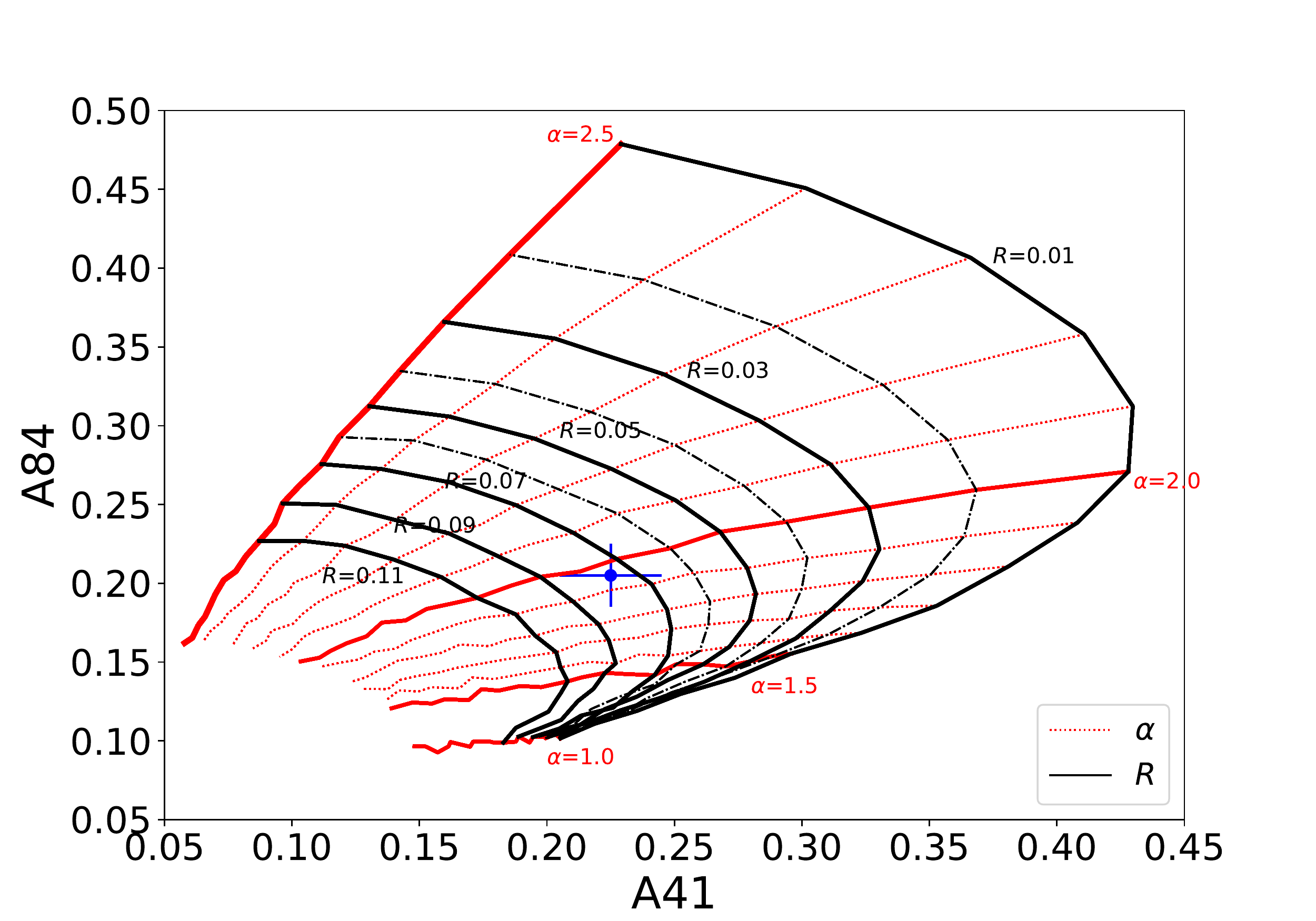}
\caption{
Theoretical [A84,A41] values measured from the accretion disc H$\alpha$
emission-line profile for each [$\alpha,R$] pair \citep{Smak81}. The  red  
and black lines represent constant $\alpha$ and $R$, 
respectively. Our observed H$\alpha$ value is shown as the solid filled 
blue circle.
}  
\label{fig10:fig_smak}
\end{figure}

\begin{figure*}
\centering
\includegraphics[width=0.9\linewidth]{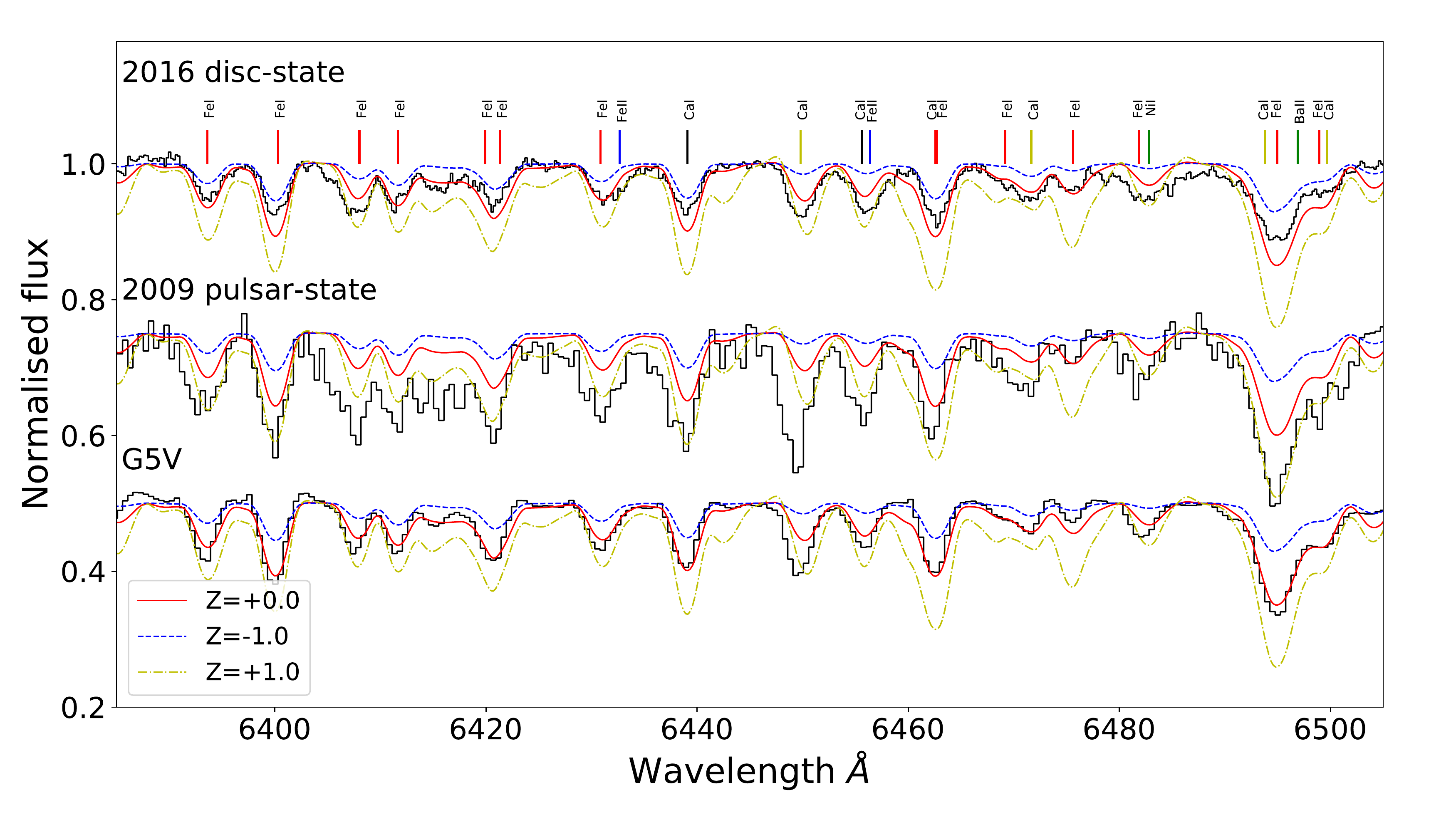}
\caption{
A comparison of the \target\ pulsar- and disc-state spectrum with synthetic 
spectra. From top to bottom: the 2016 disc-state, 2009 pulsar-state 
spectrum and the spectrum of a rotationally-broadened G5V star. 
We also over-plot the rotationally broadened synthetic spectra of a 5600\,K 
$\log\,g$=4.5 star with Z = $-$1.0 (blue dashed line), $+$0.0 (red solid line) 
and $+$1.0 (yellow dot-dashed line). The observed spectra have been shifted 
to the rest frame of the synthetic spectrum.
}  
\label{fig11:fig_spectrum_Z}
\end{figure*}

\subsection{Accretion disc size}

One can estimate the inner and outer accretion disc radii in accreting 
MSPs by measuring the emission-line wing and peak separations, respectively. 
The Keplerian velocity of an annulus of gas in Keplerian motion at radius $r$ 
around a neutron star of mass $M_{\rm 1}$ orbiting at an inclination angle $i$ 
is given by

\begin{equation}
V_{\rm kep} = \sqrt{ \frac{ G M_{\rm 1} }{ r } } \sin i.
\end{equation}

\noindent
For a double-peaked emission-line profile from an accretion disc, the 
Keplerian velocity of the outer disc radius ($r_{\rm out}$) corresponds to 
half the peak-to-peak separation, while the Keplerian velocity at the 
inner accretion disc radius ($r_{\rm in}$) corresponds to half the wing 
separation (full width at zero intensity). We measure velocities using 
spectra in which the wings and peaks of the accretion disc emission-line 
are not affected by bright-spot emission or absorption from the secondary 
star, which is around orbital phase 0.0. We use the average 2016 X-SHOOTER 
disc-spectra taken around orbital phase 0.0 and measure the H$\alpha$ half 
peak-to-peak separation and half wing separation to be 367$\pm$6\kms and 
$\sim$2000\kms, respectively. Note that the wing separation is difficult 
to establish because of the uncertainty in where the line wings end and 
the continuum begins and possible contamination from other absorption and 
disc emission lines. We estimate the ratio $R=r_{\rm in}/r_{\rm out}$ to 
be $\sim$0.03. However, it should be noted that there is clear evidence 
from studies of cataclysmic variable that the for the outer disc is 
sub-Keplerian \citep{Wade88} and so $R$ is underestimated. Accounting for 
this $\sim$20 per cent effect, gives $R\sim$0.04

We can also estimate $R$ using the method developed by \citet{Smak81}. 
Smak's method assumes an axially symmetric accretion disc in Keplerian 
motion and a power-law flux distribution $f \propto r^{-\alpha}$ for the 
disc emission. He found a relationship between 
the pairs [A84,A41] and [$\alpha,R$], where parameters 
A84 = $\log W_{\rm 0.4} - \log W_{\rm 0.8}$ and 
A41 = $\log W_{\rm 0.1} - \log W_{\rm 0.4}$ can be measured from the 
emission-line profile ($W_{\rm 0.8}$, $W_{\rm 0.4}$ and $W_{\rm 
0.1}$ are the emission-line widths at the fractions 0.8, 0.4 and 0.1 of 
the peak height above the continuum, respectively). We apply Smak's method 
to the average 2016 X-SHOOTER H$\alpha$ disc-state spectrum taken around 
orbital phase 0.0.  In Fig.\,\ref{fig10:fig_smak} we show the theoretical 
[A84,A41] pairs for a given [$\alpha,R$] as well as our observed values, 
[A84,A41] = [0.19,0.22], which corresponds to $R\sim$0.075 and $\alpha 
\sim$2.0.

Assuming that the accretion disc radius does not exceed the maximum disc 
radius determined by tidal interactions with the secondary star, 
approximated by $r_{\rm tidal}/a$ = 0.6 /(1 + q) $\sim$0.94 $\rm R_{\sun}$ 
($q$=0.141, $M_{\rm 1}$+$M_{\rm 2}$=2.0 \msun\ and $a$ is the binary 
separation; \citealt{Warner03}), and $R\sim$0.040--0.075, we estimate 
$r_{\rm in}\sim$0.04--0.07\,$\rm R_{\sun} \sim$ 26,000--50,000\,km. This 
is the location of the inner disc radius that produces the bulk of the 
H$\alpha$ emission which lies further out from from the inner edge of the 
accretion disc $\sim150$\,km \citep{Archibald09,Deller12}.

\citet{Wang09} also estimate $r_{\rm in}$ by modelling the 2001 disc-state continuum 
spectrum with a simple accretion disc model. They obtained $r_{\rm in}$ $\sim$14,000\,km. 
which is about a factor of 2 closer to the neutron star compared to what we measure 
from our 2016 disc-state spectrum. However, it should be noted  that
the 2001 spectrum \citep{Wang09} was taken just before \target\ switched to the pulsar-state 
sometime between 2001--2003 \citep{Woudt04}, whereas our 2016 spectrum was just after \target\ 
transitioned into the disc-state in 2013 \citep{Stappers14}. This suggests 
that during the extended period of accretion in the disc-state, the region responsible for the H$\alpha$ emission 
drifts inwards towards the neutron star, before transitioning to the pulsar-state.

\subsection{A peculiar secondary star}

In Fig.\,\ref{fig11:fig_spectrum_Z} we show the variance-weighted 
Doppler-averaged spectrum of \target\ taken in 2009 and 2016, as well a 
G5V template star artificially broadened by 75\kms\ for comparison.  As 
one can see, the pulsar-state 2009 spectrum shows deeper absorption lines 
compared to the F8 or G5V star, in particular the Ca\,\textsc{i} lines at 
6439.075\ang, 6493.781\ang\ and Ca\,\textsc{i} 6462.567\ang, and the 
$\sim$6495\ang\ blend of Ca\,\textsc{i} 6493.781\ang, Fe\,\textsc{i} 
6494.980\ang\ and Ba\,\textsc{ii} 6496.897\ang. This can also be seen in 
the value for $f_{\rm star}$ obtained in Section\,\ref{sec:analysis}, 
which are all greater than 1 for all spectral type. A full chemical 
analysis is beyond the scope of this paper, but we compare the observed 
\target\ spectrum with synthetic high-resolution spectra from the PHOENIX 
library \citep{Husser13}. We show the synthetic spectrum for a 5600\,K 
$\log\,g$=4.5 star with a metallicity Z of $-$1.0, 0.0 and $+$1.0. We find 
that the 2016 disc-state spectrum of \target\ and the G5V spectrum are 
reasonably well described by the Z=0.0 synthetic spectrum. However, the 
2009 pulsar-state spectrum is better described by the Z=$+$1.0 synthetic 
spectrum. Fe and Ca seems to be over-abundant in the atmosphere of the 
secondary star, and so makes the secondary star in \target\ peculiar. In 
general the features in the 2009 pulsar-state spectrum are $\sim$5 times 
stronger than in the 2016 disc-state spectrum.

\subsection{Spectral type changes along the orbit}

The light curves taken when \target\ was in the pulsar-state 
\citep{Woudt04, TA05} show an asymmetric single-humped modulation, which is 
expected from the combined effects of the tidally-locked secondary 
star's ellipsoidal modulation and the high-energy emission from the pulsar 
wind heating of the inner face of the secondary star. The light curve 
observed in the disc-state is very similar in shape and amplitude 
\citep{Kennedy18}, suggesting that the effect of heating on the secondary 
star is the same when the system is in the pulsar- or disc-state and that 
the main source of irradiation of the secondary star is the high-energy 
emission from the pulsar relativistic wind. The asymmetric maximum 
suggests that the heating does not come directly from the isotropic pulsar 
wind, but is due to non-thermal X-ray emission produced by the 
intra-binary shock between the pulsar and secondary star's wind 
\citep{Romani16}.

Given that heating with a bolometric irradiating luminosity of 
$10^{34}$\erg\ \citep{MC15} has a pronounced effect on the phase-resolved 
$V$-band light curves in the pulsar-state, we expect to see similar 
effects on the phase-resolved spectroscopy. Indeed, the 2016 disc-state 
spectra show evidence for a spectral type change across the orbit, from G5 
at phase 0.0 to F6 at phase 0.5, corresponding to a temperature change of 
5650\,K to 6340\,K, respectively \citep{Pecaut13}. This is also supported 
by the significant eccentric fit to the metallic absorption-line radial 
velocity curve (see Table\,\ref{table:rv}), where irradiation distorts the 
radial velocity curve of the secondary star, as the centre-of-light of the 
absorption lines changes with orbital phase. Significant eccentric orbits 
are not detected in the 2009 pulsar-state and 2014 disc-state spectra, 
because of the lower quality and spectral resolution of the 2009 and 2014 
spectra compared to the 2016 disc-state spectra. The moderate inclination 
angle of 54$^\circ$ of the system \citep{MC15} implies that the contrast 
between the observed spectral type at different orbital phases is reduced, 
which means that the observer always sees the heated inner face of the 
secondary star and thus the true spectral type of the secondary star is 
cooler. Simultaneous modelling of the optical light- and radial-velocity 
curves during the pulsar-state should reveal the true spectral type of the 
secondary star \citep{Shahbaz17,Linares18}

\begin{figure}
\centering
\includegraphics[width=1.0\linewidth]{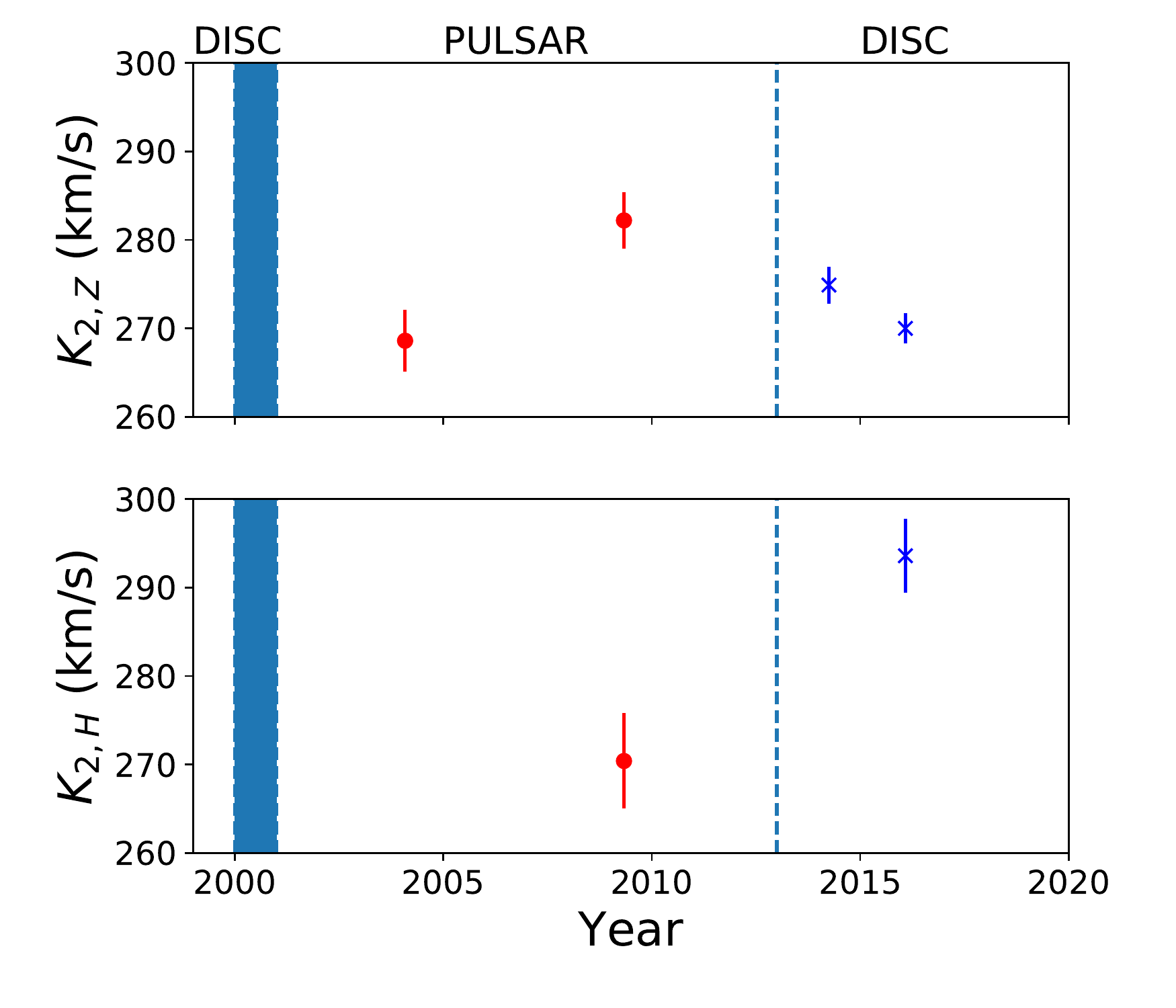}
\caption{
The observed secondary star's metallic (top) and H$\alpha$ (bottom) 
absorption-line radial velocity semi-amplitude using optical spectra taken 
in 2004 and 2009 in the pulsar-state (filled red circles) and in 2014 and 
2016 in the disc-state (blue crosses), respectively (see 
Table\,\ref{table:rv}). The vertical band marks the time when the 
system transitioned from a LMXB to a MSP around 2000-2001 \citep{Wang09} 
and from a MSP to a LMXB in 2013 \citep{Stappers14}.
}  
\label{fig12:rv_compare}
\end{figure}

\subsection{The secondary star's radial velocity curve}

The absorption lines arising from the secondary star are Doppler shifted 
as the star moves in its orbit. Therefore, the radial velocity at a given 
orbital phase depends on the observed centre-of-light of the lines that 
are used to determine the radial velocities. Heating shifts the 
centre-of-light of the secondary, weighted by the strength of the 
absorption lines, from the centre-of-mass of the star. This results in a 
distortion of the radial velocity curve leading to an apparent 
elliptical/eccentric orbit with a spurious radial velocity semi-amplitude 
\citep{Davey92,Shahbaz00}. Variable amounts of heating will produce 
different radial velocity semi-amplitudes.

In Fig.\,\ref{fig12:rv_compare} we show the different values for $K_{\rm 
2,Z}$ and $K_{\rm 2,H\alpha}$ determined between 2004 and 2016, when the 
system was either in the pulsar- (P) or disc- (D) state. The value for 
$K_{\rm 2,Z}^{\rm P}$ determined in 2004 and 2009 in the pulsar-state are 
different at the 2.8$\sigma$ level, and so we can assume that there is no 
significant change in $K_{\rm 2,Z}$ The value for $K_{\rm 2,Z}^{\rm D}$ 
determined from the 2014 and 2016 spectra, when the system was in the 
disc-state are similar at the 1.8$\sigma$ level, which is not surprising 
given that the X-ray and $\gamma$-ray flux are also very similar (see 
Fig.\,\ref{fig1:lat}). During the 2009 to 2014 transition the change in 
$K_{\rm 2,Z}$ is only significant at the 1.9$\sigma$ level. 
However, the 2009 to 2016 transition the change in 
$K_{\rm 2,Z}$ is significant at the 3.4$\sigma$ level which can be explained by 
the accretion disc shadowing (see Section\,\ref{sec:ptod_rv}).


\begin{figure*}
\subfloat[Radial velocity curve deviation and projected maps. ]
{\includegraphics[height=70mm]{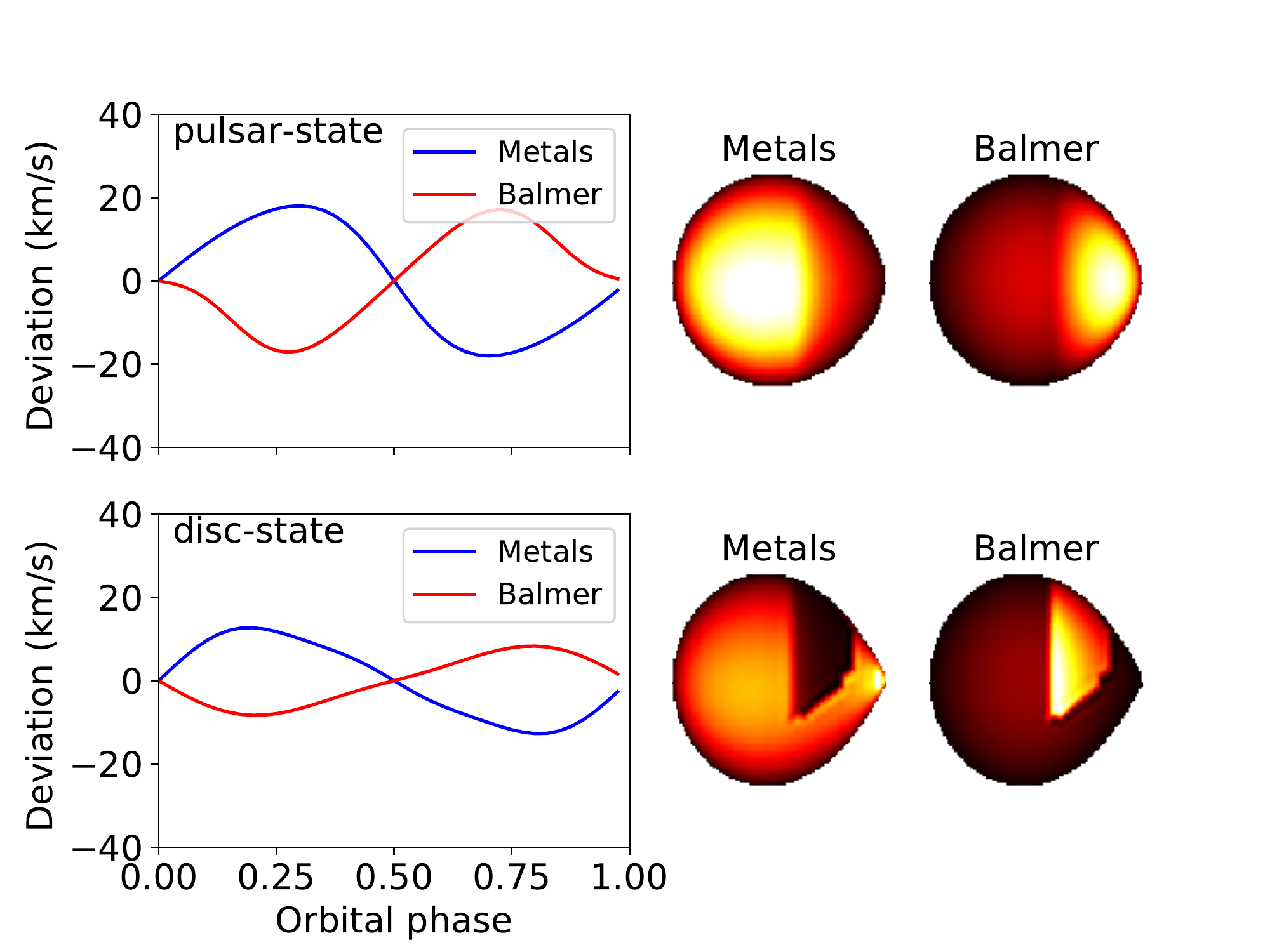}
\label{fig13:map_profile_a}}
\subfloat[Simulated absorption-line profile.]
{\includegraphics[height=65mm]{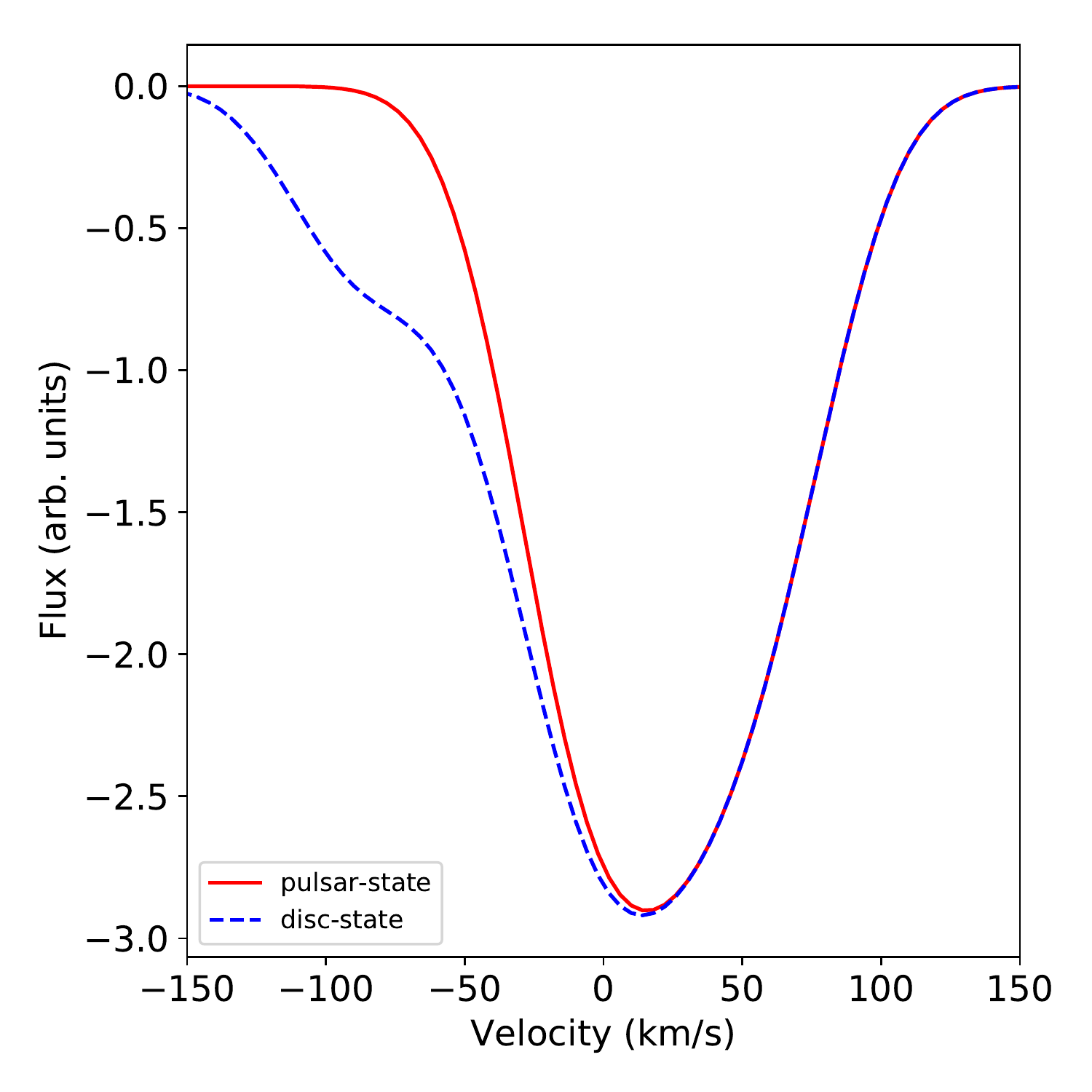}
\label{fig13:map_profile_b}}
\caption{
(a) We show the apparent radial velocity curve deviation from a circular orbit for 
the pulsar-state (top panel) and disc-state (bottom panel). In each panel 
we plot the radial velocity deviation using the metallic absorption lines 
in the 6000--6500\AA\ spectral range (solid blue) and the Balmer 
absorption line (dashed red). We assume a $T_{\rm eff}$=5660\,K secondary 
star, $q$=0.14, $i$=43$^\circ$, $K_{\rm 2}$=270\kms\ and $L_{\rm 
X}=10^{34}$\erg. For the pulsar-state we assume $f_{\rm Roche}$=0.83, with 
no accretion disc and for the disc-state we assume $f_{\rm Roche}$=1 and 
an accretion disc that extends to the tidal radius with an opening angle 
of 20$^\circ$. On the right we show projected maps of the observed 
metallic and Balmer line-strength distribution on the secondary star at 
orbital phase 0.75 using \textsc{xrbcurve} \citep{Shahbaz03a, Shahbaz17}.
(b) We show the simulated absorption-line profile in the pulsar-state (red 
solid line) and disc-state (blue dashed line), observed at orbital phase 
0.25.  The same model parameters as above are used. Note the change in the 
line profile shape.
}  
\end{figure*}


\subsubsection{The pulsar-state radial velocity curve}
\label{sec:pulsar_rv}

We can use our X-ray binary model \textsc{xrbcurve} \citep{Shahbaz03a, 
Shahbaz17} to determine the line flux distribution and hence radial 
velocity curve arising from the secondary star in \target, for a system in 
the pulsar- and disc-state, where one may expect different amounts of 
heating. An important parameter that determines the shape and amplitude of 
the radial velocity curve is the effects of heating and how the 
line-strength versus temperature relation used in determining the radial 
velocity. We assume the ''deep heating" approximation which means that 
each element radiates as predicted by a model atmosphere for a single 
star. The spectral type of the secondary stars in MSPs and X-ray binaries 
are typically later than F and their spectra contain metal absorption 
lines. The strongest absorption metallic lines in the red part of the 
optical spectrum are the Fe\,\textsc{\i}, Ca\,\textsc{i} and 
Ba\,\textsc{ii} lines (see Fig.\,\ref{fig11:fig_spectrum_Z}). Using UVESPOP 
spectra we find that the line strength versus temperature relation of the 
metallic absorption lines in the 6000--6500\ang\ spectral range decreases 
with increasing temperature. Similarly, for the H$\alpha$ absorption-line 
(Balmer), we find that the line strength increases with increasing 
temperature \citep{Shahbaz17,Linares18}. Therefore, we can compute the 
secondary star's metallic ($K_{\rm 2,Z}$) or Balmer ($K_{\rm 2,H}$) 
absorption-line radial velocity semi-amplitude. 
Equations\,\ref{eqn:c1} to \ref{eqn:c5} are the conditions
we find due the effects of irradiation and/or accretion disc on the 
metallic and Balmer line radial velocity curves semi-amplitudes. 

\begin{subequations}
\begin{align}
K_{\rm 2,Z}^{\rm P} > K_{\rm 2,H}^{\rm P}     \label{eqn:c1} \\
K_{\rm 2,Z}^{\rm D} > K_{\rm 2,H}^{\rm D}     \label{eqn:c2} \\
K_{\rm 2,Z}^{\rm D} < K_{\rm 2,Z}^{\rm P}     \label{eqn:c3} \\
K_{\rm 2,H}^{\rm D} > K_{\rm 2,H}^{\rm P}     \label{eqn:c4} \\
K_{\rm 2,Z}^{\rm D} > K_{\rm 2,H}^{\rm D}     \label{eqn:c5} 
\end{align}
\end{subequations}

\noindent
In the pulsar-state, high energy relativistic particles produced by the 
pulsar wind are mostly likely responsible for heating the secondary star's 
inner face. The hot inner face thus produces less metallic line flux 
compared to the non-heated face of the star, and so the centre-of-light is 
shifted towards the non-heated side of the star, resulting in a spuriously 
high radial velocity semi-amplitude;  $K_{\rm 2,Z}^{\rm P} > K_{\rm 2}$. 
In Fig.\,\ref{fig13:map_profile_a} we show the metallic and Balmer 
absorption-line radial velocity curve deviations. The model predicts that 
the metallic radial velocity semi-amplitude is greater than the Balmer 
radial velocity semi-amplitude in both the  pulsar- and disc-states
(see Equation\,\ref{eqn:c1}). 
In contrast, the hot inner face produces more Balmer flux on the heated 
face compared to the non-heated face of the star, and so the 
centre-of-light is shifted towards the heated side of the star, resulting 
in a spuriously low radial velocity semi-amplitude;  $K_{\rm 2,H}^{\rm P} 
< K_{\rm 2}$.  Our ISIS 2009 observations in the pulsar-state 
imply 270.4 $<K_{\rm 2}<$ 282.2\kms\ (or $K_{\rm 2}$ = 276.3$\pm$5.6 \kms).

\subsubsection{The pulsar- to disc-state radial velocity}
\label{sec:ptod_rv}

The pulsar- to disc-state transition involves an increase in the gamma-ray 
and X-ray flux \citep{Takata14}, resulting in a change in the 
metallic and Balmer line radial velocity 
semi-amplitude because of the effects of the accretion disc shadowing. In 
Fig.\,\ref{fig13:map_profile_b} we show the effects the accretion disc has on 
the shape of the metallic absorption-line profile. We show the simulated 
line profile in the pulsar- and disc-state observed at orbital 
phase 0.25. The elements of area near the inner Lagrangian point  
contribute flux to the negative wing of the line profile. We assume a 
bolometric irradiating luminosity of 10$^{34}$\erg\ which is the same in 
the pulsar- and disc-state. In the pulsar-state elements of area on the 
hot irradiated inner face of the star produce less metallic line flux 
(because of the line-strength versus temperature relation)  compared to 
the non-heated face of the star, and so the centre-of-light of the line 
profile (blue solid line) is shifted towards the non-heated side of the 
star. The appearance of an accretion disc shadows the elements area near 
the inner Lagrangian point (negative wing of the line profile at phase 
0.25), which intrinsically have lower temperatures because of gravity 
darkening, and so contribute more flux than in the pulsar-state. The line 
profile is skewed and the effect of the disc shadowing is to shift the 
centre-of-light of the line profile towards the inner Lagrangian point, 
resulting in a lower radial velocity semi-amplitude compared to in the 
pulsar-state (see Equation\,\ref{eqn:c3}). For the 
H$\alpha$ absorption-line, the opposite line-strength versus temperature 
relation implies that the centre-of-light of the absorption-line profile 
is shifted away from the inner Lagrangian point. This results in a higher 
semi-amplitude compared to when there is no accretion disc 
(see Equation\,\ref{eqn:c4}). Finally the accretion disc shadowing also 
gives the condition in Equation\,\ref{eqn:c5}.

We find that the 2009 pulsar-state data satisfies equation\,\ref{eqn:c1} and 
the metallic and Balmer radial velocities determined 
from the 2009 to 2014/2016 transition satisfies
equation\,\ref{eqn:c3} and \ref{eqn:c4}, respectively.
However, the 2016 disc-state data does not satisfy equation\,\ref{eqn:c2} or 
\ref{eqn:c5}, but this could be due to problems in 
disentangling the narrow H$\alpha$ absorption-line modulation from the 
double-peaked disc emission-line profile. 
Therefore the observed changes in the $K_{\rm 2,Z}$ and 
$K_{\rm 2,H}$ in the pulsar- and disc-state can be explained to some 
extent by the presence of an accretion disc.

\subsection{The Roche lobe filling factor in the pulsar-state}

In contrast to the disc-state, in the pulsar-state no accretion disc is 
present and hence the secondary star does not fully fill its Roche lobe. Indeed, 
observations of the MSP secondary stars in the pulsar-state show that they 
underfill their Roche lobes; the Roche lobe filling factor is in the range 
$f_{\rm Roche}$=0.80--0.95 \citep{Shahbaz17, Linares18}.  Given that the 
star's rotational broadening depends on its radius, the star's radius is 
smaller in the pulsar-state compared to in the disc-state. It can be shown 
that the star's equivalent volume radius depends on the binary mass ratio 
and Roche lobe filling factor, as well as its radial velocity 
semi-amplitude \citep{Shahbaz17}. Using equation (2) in \citet{Shahbaz17} 
we can calculate the change in the rotational broadening for different 
values of $f_{\rm Roche}$ ($q$=0.14, see Section\,\ref{sec:ptod_rv}) and 
compare it to the observed change in \vsini. Given the observed 
X-ray/gamma-ray luminosities, the effects of irradiation in the 
determination of \vsini\ in the two states are similar and so 
appropriately cancel out. In Section\,\ref{sec:results} we determine the 
secondary star's rotational broadening in the pulsar-state to be larger 
than in the disc-state (see Table\,\ref{table:rv}). However, it should be 
noted that the pulsar-state \vsini\ measurement has a large error. The 
ratio of pulsar-state to disc-state \vsini\ value is 1.037$\pm$0.038. 
Using the 3$\sigma$ lower limits, we find a change in \vsini\ of $>$0.923, 
which corresponds to a Roche lobe filling factor of $f_{\rm Roche}$>0.78, 
agreeing with what was obtained by \citet{MC15} by modelling the $V$-band 
light curve obtained by \citet{TA05} in the pulsar-state.

\subsection{Irradiating luminosity}
\label{sec:Lirr}

In the pulsar- and or disc-state, the relativistic pulsar wind and and/or 
X-ray/gamma-ray emission from the accretion disc are the dominant sources 
of irradiation. We can compute the irradiating luminosity temperature 
difference of hemispheres of the secondary star and compare it to the 
observed X-ray and gamma-ray luminosities, as well as to the pulsar's spin 
down luminosity ($L_{\rm sd}$) and determine the source of the driving 
mechanism. The ``irradiation temperature'' is given by $T_{\rm irr}^4 = 
T_{\rm day}^4 -T_{\rm night}^4$, where $T_{\rm night}$ and $T_{\rm day}$ 
are the temperatures observed at orbital phase 0.0 and 0.5, respectively. 
The irradiating luminosity is then $L_{\rm irr}=4\pi a^2\sigma T_{\rm 
irr}^4$ where $a$ is the orbital separation, which is related to $L_{\rm 
sd}$ via the efficiency of irradiation parameter; $\eta_{\rm irr}=L_{\rm 
irr}/L_{\rm sd}$. Using $T_{\rm night}$=5650\,K and $T_{\rm day}$=6340\,K 
\citep{Pecaut13} we find $T_{\rm irr}$=4943\,K and $L_{\rm 
irr}=6.5\times10^{33}$ \erg, which implies $\eta_{\rm irr}$=14 percent. 
For \target\ in the disc-state, the observed X-ray (0.5--10 keV) 
luminosity is $2\times10^{33}$ \erg\ \citep{Takata14}, the gamma-ray 
(0.2--20 GeV) luminosity is $6\times10^{33}$ \erg\ \citep{Stappers14} and 
the pulsar's spin-down luminosity (corrected for the Shklovskii effect) is 
$L_{\rm sd}=4.8\times10^{34}$ \erg\ \citep{Strader18}.
The irradiation efficiency of 14 percent we derive is similar to what is 
observed in other MSPs \cite{Breton13}. Indeed, the energetics suggest 
that the pulsar's the relativistic wind, powered by the rotational spin 
down of the neutron star can drive the observed heating mechanism in the 
disc-state. However, in \target\ the observed X-ray and gamma-ray 
(most likely due to  accretion)
luminosities are also sufficient to provide the observed irradiating 
luminosity.

\subsection{The system parameters}
\label{sec:sys}

\citet{Archibald09} determined the binary mass ratio of 
$q$=0.141$\pm$0.002 ($q$=$K_{\rm 1}$/$K_{\rm 2}$) by measuring the radial 
velocity semi-amplitude of the neutron star determined from radio timing 
$K_{\rm 1}$=38 \kms and the secondary star's radial velocity 
semi-amplitude 268$\pm$4 \kms determined by \citep{TA05}. We update this 
value using our $K_{\rm 2}$ value determined in 
Section\,\ref{sec:pulsar_rv} to obtain $q$=0.137$\pm$0.003. We can also 
determine the mass ratio from the fact that the Roche lobe filling star's 
rotational broadening radius depends only $q$ and $K_{\rm 2}$ in the 
disc-state \citep{Wade88,Shahbaz17}. Using \vsini=74.9$\pm$0.9\kms\ and 
our $K_{\rm 2}$ value we obtain $q$=0.20$\pm$0.03. It should be noted that 
this value is uncertain because the limb-darkening coefficient used in the 
determination of \vsini\ leads to an overestimation of secondary star's 
rotational broadening. Also, irradiation leads to an underestimation of 
\vsini\ \citep[see][and references within]{Shahbaz03b}. Despite this, the 
value for $q$ is consistent, at the 2-$\sigma$ level, with the revised value 
determined earlier.

From the mass function equation, one can use $K_{\rm 2}$, $q$, the orbital period 
and the binary inclination $i$ to determine the masses of the compact object 
and secondary star ($M_{\rm 2}$). Using our values for  
$K_{\rm 2}$ = 276.3$\pm$5.6 \kms and $q$=0.137$\pm$0.003 we obtain

\begin{equation}
M_{\rm 1} = \frac{0.56 \pm 0.03}{\sin^3\,i} M_{\odot}, \hspace{15mm}
M_{\rm 2} = \frac{0.077 \pm 0.005}{\sin^3\,i} M_{\odot}  
\end{equation}

\noindent
For a neutron star with a mass in the range 1.4 to 3 \msun, corresponding 
to the canonical and maximum theoretical mass of a neutron star 
\citep{Chamel13}, respectively, implies $i$ between 47 and 35$^\circ$, 
respectively.

\cite{TA05} modelled the multi-colour ($B$, $V$ and $I$) optical 
lightcurves of \target\ in the 2009 pulsar-state with a fully Roche-lobe 
filling irradiated secondary star model. They noted that assuming that if 
the secondary fills its Roche lobe its mass can be inferred from the 
distance ($D_{\rm kpc}$), because the distance determines the temperature 
and size of the star. If the star underfills its Roche lobe then $D_{\rm 
kpc} < 2.20 (M_{\rm 2}/M_{\odot} )^{1/3}$. \citet{Deller12} used this 
relation with $K_{\rm 2}$ measured by \cite{TA05}, $K_{\rm 1}$ from radio 
timing and the distance $D_{\rm kpc}$ of 1.37 kpc 
 from the known radio parallax to 
determine $M_{\rm 1}$. We use the same method but with our revised values 
for $K_{\rm 2}$ and $q$ to obtain $M_{\rm 1}>1.76\pm$0.16 \msun, $M_{\rm 
2}>0.24\pm$0.02\msun\ and $i$<43$\pm$2$^\circ$. The mass estimates are 
lower limits because the secondary star in the pulsar-state can 
substantially underfills its Roche lobe.

The $V$-band optical lightcurve and a radial velocity curve presented in 
\cite{TA05} was also modelled by \cite{MC15}. Allowing for the secondary 
star's Roche lobe filling factor to vary, they find that the secondary 
star underfills its Roche lobe and $i\sim$54$\pm$5$^\circ$. This inclination angle is much higher than what we 
estimate above and gives an unusually low mass for the neutron star; 
$M_{\rm 1}=1.1\pm$0.2 \msun. Given that we expect relatively massive ($\sim$1.8\msun) neutron 
stars to reside in MSPs, \citep{Strader18} we suspect their determination of 
$i$ is biased. Simultaneous modelling of multi-band lightcurves with high 
resolution spectroscopy is needed. This will allow one to determine 
effective temperature of the secondary, its true radial velocity 
semi-amplitude, inclination angle and hence the mass of the neutron star.

\section*{Conclusions}

We present time-resolved optical spectroscopy of the binary millisecond 
pulsar PSR\,J1023+0038 during its 2009 radio-powered pulsar-state and 
during its accretion-powered disc-states in 2014 and 2016. Below we list 
the main results of this paper.

\begin{enumerate}

\item
We provide observational support for the companion star being 
heated during the disc-state. We observe a spectral type change along the 
orbit, from $\sim$G5 to $\sim$F6 at the secondary star's superior and 
inferior conjunction, respectively, which correspond to the 
"day" and "night" side temperatures of the secondary star.
We find that the irradiating 
luminosity can be powered by the spin down luminosity of the neutron star, 
as is the case on many other MSPs, or by the accretion luminosity of the 
accretion disc.

\item
We determine the secondary star's radial velocity semi-amplitude from the 
metallic (primarily Fe and Ca) and H$\alpha$ absorption lines during the 
pulsar- and different disc-states. We find that the observed changes in 
the metallic radial velocity semi-amplitude is only significant (at the 3.4$\sigma$ level)
for the 2009 pulsar-state to 2016 disc-state transition, which can be explained by
the accretion disc shadowing.
The metallic and H$\alpha$ radial velocity semi-amplitude determined from 
the 2009 pulsar-state observations allows us to constrain the secondary 
star's true radial velocity $K_{\rm 2}$=276.3$\pm$5.6 \kms\ and the binary 
mass ratio $q$=0.137$\pm$0.003.

\item
Doppler maps of the disc-state spectra show characteristic double-peaked 
emission-line profile arising from the accretion disc, a narrow 
absorption, which is in phase with the secondary star, and a narrow 
emission-line feature which is shifted by $\sim$-0.1 phase with respect to 
the secondary star, which arises from the accretion disc.
From the average emission-line profile of the accretion disc in 2016, we 
place constraints on the inner to outer disc radii ratio, 0.04--0.075.

\item
By comparing the observed metallic and H$\alpha$ absorption-line radial 
velocity semi-amplitudes with model predictions, we can explain the 
observed semi-amplitude changes during the different pulsar states and 
during the pulsar/disc-state transition as being due to different amounts 
of heating and the presence of an accretion disc, respectively.

\end{enumerate}

\section*{ACKNOWLEDGEMENTS}

We thank Tom Marsh for the use of his \textsc{molly}, \textsc{pamela} and 
Doppler tomography programs. We acknowledge the use of data from the UVES 
Paranal Observatory Project (ESO DDT Program ID 266.D-5655). M.L. is 
supported by EU's Horizon 2020 programme through a Marie Sklodowska-Curie 
Fellowship (grant No. 702638). Based on observations made with the WHT 
telescope operated by the Instituto de Astrof\'\i{}sica de Canarias in the 
Spanish Observatories of el Roque de los Muchachos (La Palma). Based on 
observations made with ESO Telescopes at the La Silla Paranal Observatory 
under ESO programme 096.D-0808. This paper makes use of data obtained from 
the Isaac Newton Group Archive which is maintained as part of the CASU 
Astronomical Data Centre at the Institute of Astronomy, Cambridge. The 
Starlink software \citep{Currie14} is currently supported by the East 
Asian Observatory.

\medskip

\noindent
{\it Facilities:} WHT (ISIS), WHT (ACAM), VLT (X-SHOOTER)


\end{document}